\PassOptionsToPackage{hyphens}{url}
\documentclass[pageno]{jpaper}

\pretitle{\begin{center}
\normalfont\Large\bfseries}

\usepackage[dvipsnames]{xcolor}
\usepackage[normalem]{ulem}
\usepackage[ruled,linesnumbered]{algorithm2e}
\usepackage[utf8]{inputenc}
\usepackage{framed}
\usepackage{multirow}
\usepackage{color}
\definecolor{shadecolor}{rgb}{0.92,0.92,0.92} 
\usepackage{listings}
\usepackage{verbatim}
\usepackage[frozencache=true,cachedir=_minted-output]{minted}
\usepackage{tikz}
\usepackage{appendix}
\usepackage{graphicx}
\usepackage{float}
\usepackage{subfig}
\usepackage{environ}
\usepackage{comment}
\usepackage{booktabs}
\usepackage{ragged2e}
\usepackage{bm}
\usepackage{amsbsy}

\usepackage{amssymb}%
\usepackage{pifont}%
\newcommand{\cmark}{\textcolor{JungleGreen}{\ding{51}}}%
\newcommand{\xmark}{\ding{55}}%

\newcommand*\circled[1]{\tikz[baseline=(char.base)]{
            \node[shape=circle,draw, fill=black, text=white, inner sep=0.1mm] (char) {#1};}}

\usepackage[varqu]{zi4}%
\AtBeginDocument{%
}

\makeatletter
\newcommand{\printfnsymbol}[1]{%
  \textsuperscript{\@fnsymbol{#1}}%
}
\makeatother

\author{ 
    {\normalsize Yun Chen\thanks{These authors contributed equally to this work.}}\\
    {\normalsize School of Computing}\\
    {\normalsize National University of Singapore}
  \and
    {\normalsize Lingfeng Pei\printfnsymbol{1}} \\ 
    {\normalsize School of Computing}\\
    {\normalsize National University of Singapore}
  \and
    {\normalsize Trevor E. Carlson}\\
    {\normalsize School of Computing}\\
    {\normalsize National University of Singapore}
}

\setminted[C]{fontfamily=cmtt,fontsize=\scriptsize, xleftmargin=20pt,linenos}

\newcommand{\varinum}{three}%
\newcommand{\comend}{.}%
\newcommand{\andend}{and}%

\newcommand{\name}{AfterImage}

\begin{document}

\title{Leaking Control Flow Information via the Hardware Prefetcher}

\date{}
\maketitle

\pagenumbering{gobble}

\begin{abstract}
Modern processor designs %
use a variety of microarchitectural methods to achieve high %
performance. Unfortunately, new side-channels have often been uncovered 
that exploit these enhanced designs.
One area that has received little attention from a security perspective %
is the processor's hardware prefetcher, a critical component used to mitigate DRAM latency in today's systems.
Prefetchers, like branch predictors, hold critical state related to the execution of the application, and have the potential to leak secret %
information. But up to now, there has not been a demonstration of a generic prefetcher side-channel that could be actively exploited in today's hardware.

In this paper, we present \name{}, a new side-channel that exploits the Intel Instruction Pointer-based stride prefetcher. We observe that, when the execution of the processor switches between different private domains, the prefetcher trained by one domain can be triggered in another. %
To the best of our knowledge, this work is the first to publicly demonstrate a methodology that is both algorithm-agnostic and also able to leak kernel data into userspace. %
\name{} is different from previous works, as it leaks data on the non-speculative path of execution. Because of this, a large class of work that has focused on protecting %
transient, branch-outcome-based data will be unable to block this side-channel. %
By reverse-engineering the IP-stride prefetcher in modern Intel processors, we have successfully developed three variants of \name{} to leak control flow information across %
code regions, processes and the user-kernel boundary.
We find a high level of accuracy in leaking information with our methodology (from 91\%, up to 99\%), and propose two mitigation techniques to block this side-channel, one of which can be used on hardware systems today.

\end{abstract}

\section{Introduction}
\label{sec:intro}

Modern systems leverage a variety of methods to achieve high performance, from new instructions to microarchitectural enhancements. Unfortunately, many of these techniques designed to speed up processors have also resulted in exposing new microarchitectural side-channel attacks~\cite{ge2018survey, 10.1145/3456629}. The processor cache, as an example, can significantly improve workload performance. However, its timing information can expose the memory activity of programs, which has been widely exploited as timing side-channels to leak information such as AES keys~\cite{osvik2006cache}. Recently, the branch predictor has received
significant attention for its %
ability to amount speculative execution attacks~\cite{bhattacharyya2020specrop,Kocher2018spectre, lipp2018meltdown, schwarz2019zombieload, van2020cacheout} by inducing the processor to execute mispredicted instructions. Furthermore, a number of other microarchitectural features have been newly identified as vulnerable, from TLBs and Line Fill Buffers, to a variety of other hardware structures~\cite{gras2018translation, puddu2020frontal,ren2021see,van2020cacheout}.

Hardware prefetchers have also %
been investigated as a potential side-channel and covert channel, expanding this list of vulnerable hardware structures~\cite{host,ccs}. %
Prefetchers work by preloading data into the %
cache before it is requested by the processor 
to mitigate the effects of extremely long DRAM load latencies.
In the vast majority of current Intel processors, an Instruction Pointer-based stride (IP-stride) prefetcher can be
found~\cite{whitepaper} as it is a small structure that can provide a significant performance benefit. This prefetcher learns repetitive strides between load addresses requested by the same IP. When the prefetcher obtains sufficient confidence, it will prefetch the next address into the cache, which is the sum of the current address and the previously learned stride.

Shin et al.~\cite{ccs} was the first to verify the existence of a potential data breach introduced by the Intel IP-stride prefetcher. They were able to extract the secret value of the ECDH algorithm~\cite{ECDH} in OpenSSL~\cite{OpenSSL} through a cache side-channel. While an interesting proof-of-concept, this work was only applicable to a very specific program behavior, i.e. table look-ups, and is not a general side-channel. One subsequent work~\cite{host} developed a covert channel with the use of the IP-stride prefetcher. By iteratively training and evicting the recently trained entries, the sender and the receiver were able to communicate with each other. This method was designed solely for use as a covert channel and requires both the sender and receiver to be controlled by the attacker; it cannot be extended to leak information as a side-channel. These first attempts to exploit the IP-stride prefetcher are either algorithm-specific, or can only be used as a covert channel; they cannot be used as a general-purpose side-channel for a large class of applications.

\begin{figure}[t]
    \centering
    \includegraphics[width=1\linewidth]{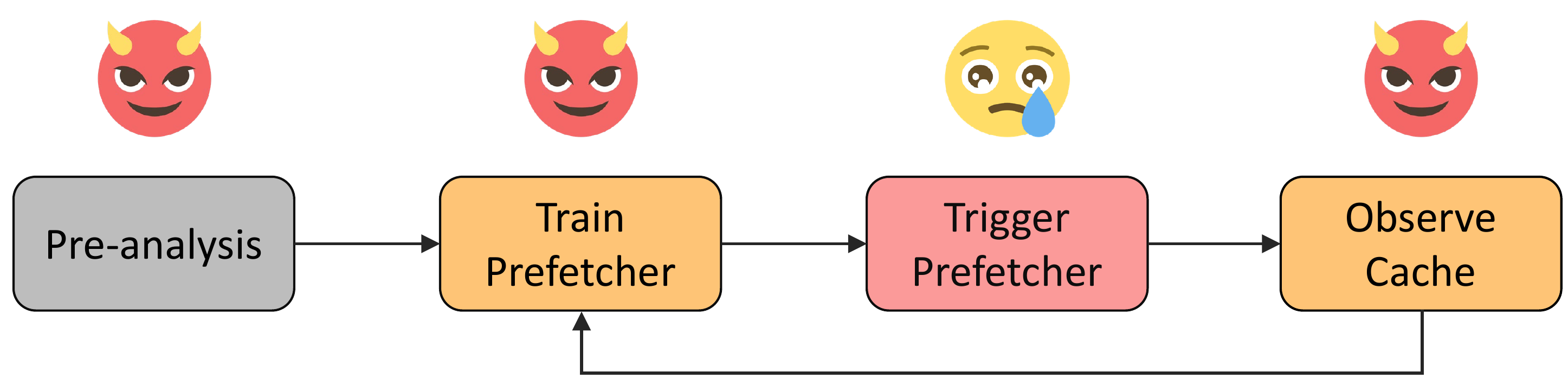}
    \caption{\name{} Overview. %
    }
    \label{fig:ow}
\end{figure}

In this paper, we demonstrate a generic hardware prefetcher-based side-channel, called \name{}, that can be exploited in today's 
processors\footnote{\name{} has been disclosed to Intel, and approved for distribution.} %
to extract secret information across program and user-kernel boundaries.
With the use of load instructions that overlap the key set of address bits of the destination IP, and a short training phase with different strides for the \texttt{if} and \texttt{else} cases of a branch, we can persuade the hardware to leak information from the next execution of an application, or from the kernel itself (See Figure~\ref{fig:ow}).

To accomplish this, we show that, when the execution of the processor switches between different %
domains, 
such as
user-kernel state switching, the IP-stride prefetcher trained by one domain can be triggered in another domain.
This provides an opportunity to leak the control flow or execution information of the application, and therefore the potential secret, even in the presence of traditional hardware isolation techniques.

In summary, we make the following major contributions:

\begin{itemize}
    \item We describe a new prefetcher side-channel, \name{}, that is able to leak the control flow of applications. The new side-channel attack does not rely on changes to the control flow graph (i.e., change the status of branch predictor) and does not rely on branch-speculative execution, defeating many recent defences against transient side-channels.
    As a side-channel attack, \name{} can achieve a high attack success rate (from 91\% to 99\%). 
    \item To accomplish the side-channel attack, we developed a custom set of microbenchmarks that revealed critical features of the Intel IP-stride prefetcher, including the number of entries, the structure of every entry, stride and confidence updating policy, indexing strategy, replacement policy, and page checking policy. By using these microbenchmarks, we present an in-depth study of the IP-stride prefetcher used in modern Intel processors. Reverse engineering critical features like the new ability %
    to prefetch past the virtual page boundary, and the fact that high-confidence matches will always trigger, have allowed us to construct this side-channel.
    \item We evaluate three attack variants
    that exploit the novel IP-stride prefetcher vulnerability: (a) cross-code region / same address-space attack, (b) cross-process attack, \andend{} (c) cross user-kernel boundary attack\comend
    
    \item We propose, implement and evaluate a light-weight mitigation strategy for future hardware designs that shows an extremely small (0.8\%) slowdown for applications which can take advantage of this prefetcher. In addition, we also propose a mitigation that can be used on today's hardware to prevent this attack. %
\end{itemize}

In the rest of this paper, we first provide a short overview of how \name{} works (Section~\ref{sec:afterimage-overview}) and provide some additional background information for this work (Section~\ref{sec:background}). Next, we study the IP-stride prefetcher present in today's hardware (Section~\ref{sec:rvp}), and discuss (Section~\ref{sec:attack}) and evaluate (Section~\ref{sec:evaluation}) the proposed side-channel and our defense options. Finally, we discuss effectiveness of existed defenses in Section~\ref{sec:discuss}, outline related work in Section~\ref{sec:related}, and conclude in Section~\ref{sec:conclusion}.

\section{\name{} Overview}
\label{sec:afterimage-overview}

The overview of \name{} is shown in Figure~\ref{fig:ow}. There are four main steps needed to leak information via this prefetcher side-channel.

    \textbf{Pre-analysis:} The attacker %
    locates %
    vulnerable code in the victim process, and generates a local version of the targeted %
    \texttt{load} instructions. %
    These \texttt{load}s masquerade as the target \texttt{load}s
    and share the same hardware entry in the prefetcher.
    
    \textbf{Train Prefetcher:} The attacker then trains the  IP-stride prefetcher locally by %
    executing
    a strided address sequence
    to 
    obtain a
    sufficient level of confidence in the prefetcher.
    
    \textbf{Trigger Prefetcher:} When the victim executes the targeted code region, the prefetcher will be automatically triggered with the previously trained stride.
    
    \textbf{Observe Cache:} The attacker detects the existence of one or more %
    strides through cache side-channels (e.g., Prime+Probe, Flush+Reload, etc.) and infers the control flow to leak the secret data. %

The flexibility of training the IP-stride prefetcher with arbitrary stride values provides a series of advantages. First, we explicitly use a stride value with a cache line granularity, which aligns with the use of traditional cache side-channels such as Prime+Probe~\cite{osvik2006cache} and Flush+Reload~\cite{yarom2014flush+}, since they can only detect information as fine as the cache line size. Moreover, we only detect whether these specific strides exist or not. This allows for a generic implementation, and requires significantly less %
effort by the attacker to extract potential secret information compared with previous works that utilize time-series analyzing methods~\cite{ccs}.
In addition, this method also improves the resolution and accuracy of \name{} compared to prior works. Our proof-of-concept is built upon the Prime+Probe and Flush+Reload cache side-channel, which sees success rate of at least 91\% and as high as 99\%.

\name{} is not susceptible to the prevailing defenses against transient execution~\cite{bhattacharyya2020specrop,koruyeh2020speccfi, loughlin2021dolma, 9251956, tol2020fastspec, yu2019speculative, qi2021spectaint}. Many of these defenses will selectively disable the speculative \texttt{load}s %
by serializing load instructions or introducing invisible speculation if the malicious code is trying to transiently access addresses that might result in data leakage. However, \name{} does not deviate from the the normal control flow or data flow of the original program and will not actively change the pipeline state of the of the program. %
These defenses are therefore currently vulnerable to the \name{} side-channel.

\section{Background}
\label{sec:background}
\subsection{Prefetching on Intel Microprocessors}
\label{sec:preback}
Prefetching is a widely adopted technique in modern processors used to mitigate the latency gap between the CPU and the memory subsystem. Prefetchers can hide the long DRAM latency by predicting and preloading data from slower memory into the high speed cache before the data is requested by the CPU. Intel processors provide both software prefetching instruction interfaces and dedicated prefetching hardware components. %
Software prefetching requires use of programmer knowledge or compiler information by inserting \texttt{PREFETCH} instructions into the program with an explicit memory address, while hardware prefetchers automatically predict the memory access address by learning the run-time memory access patterns. %
The speculation that occurs in the hardware prefetcher is different from branch predictor speculation. If the prediction of prefetching is wrong, the useless memory accesses may waste bandwidth or pollute the cache. However, the data will not be used by the processor and will not affect the normal execution of the program. %

Intel has described four hardware prefetchers %
in their processor designs~\cite{whitepaper}, and 
their features are listed in Table~\ref{tb:prefs}. The data cache unit (DCU) prefetcher, also known as next-line prefetcher~\cite{smith1978sequential}, attempts to automatically prefetch a single, subsequent %
cache line.
Data prefetch logic (DPL), i.e., the adjacent prefetcher, regards data as 128-bytes aligned blocks. A cache miss to one of the two cache lines in this block will trigger a prefetch to the pair line. The Streamer prefetcher records sequential 
positive and negative offset streams
and prefetches the next or previous several cache lines, respectively. Previous work~\cite{9229804} has shown that the Streamer will maintain the status after a context switch as well. However, it is located at the L2 cache and indexed by physical memory address. %
It will dynamically decide how many cache lines should be prefetched based on the system status (e.g., bandwidth, streaming direction, etc.). 
Therefore, the operation of these three hardware prefetchers do not have as much flexibility as the Instruction Pointer (IP)-based strided prefetcher, i.e. the IP-stride prefetcher. %

\begin{figure}[t]
    \centering
    \includegraphics[width=0.90\linewidth]{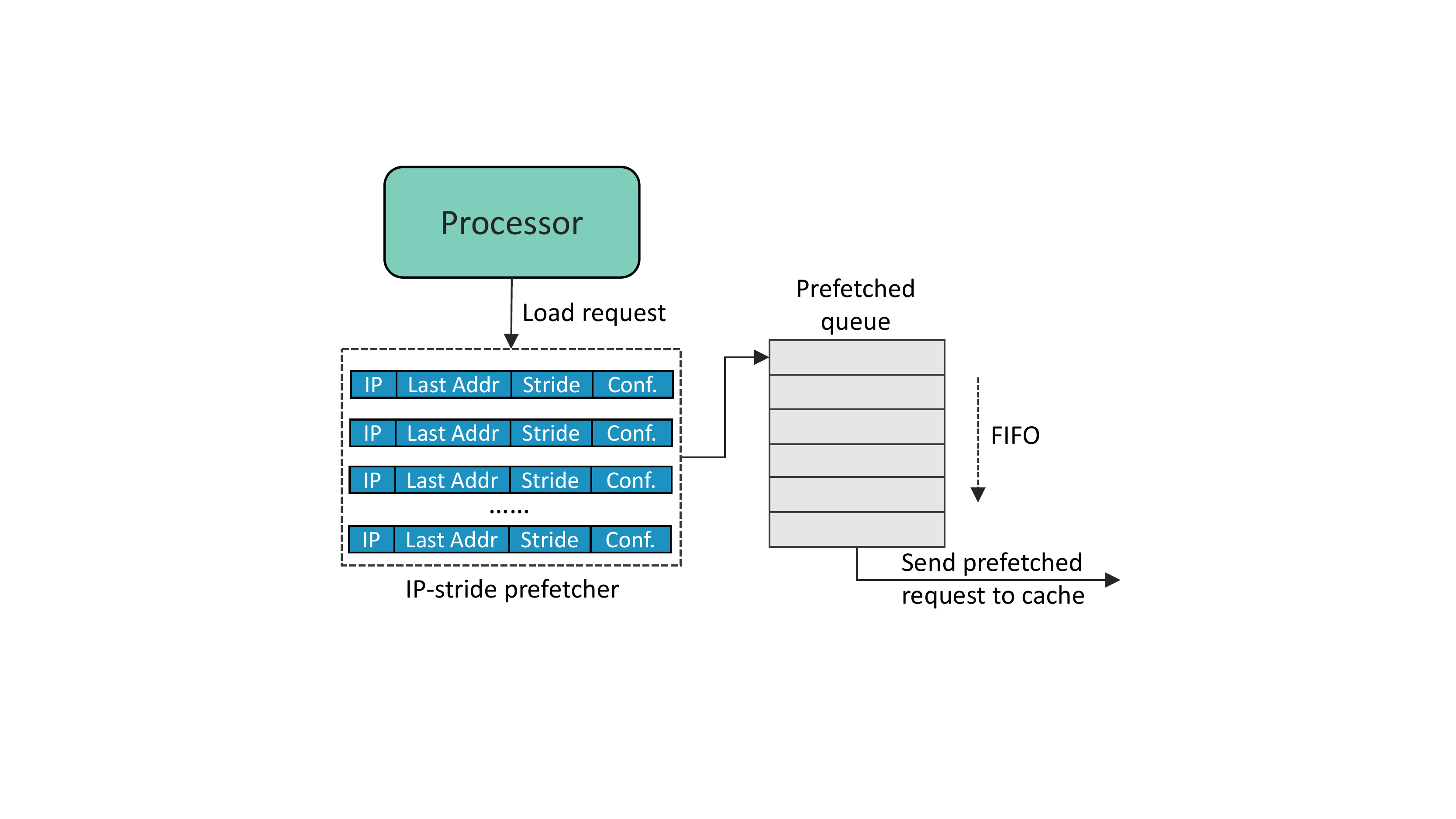}
   \caption{General architecture of the IP-stride prefetcher.  %
   }
    \label{fig:gip}
\end{figure}

The basic structure of the IP-stride prefetcher is shown in Figure \ref{fig:gip}. This prefetcher keeps track of \texttt{load} instructions with regular strides from the same IP. %
Its operation is composed of three
steps. \textbf{1. Index and Replace.} When a \texttt{load} instruction is present, it will be indexed into an entry with the same IP tag. If no such IP tag exists, a victim entry will be selected and evicted. \textbf{2. Update.} If the difference between the current address and the $Last\ Addr$ is equal to the $Stride$, the $Confidence$ will be increased, otherwise it will be decreased. However, if the $Confidence$ drops below a threshold, the $Stride$ value will be updated to a new stride as well. \textbf{3. Prefetch.} If the $Confidence$ exceeds a certain threshold, a prefetch request will be sent to the next address which is the sum of the current access address and the recorded $Stride$ value.

Although some of the parameters and structures of Intel's IP-stride prefetcher has been documented previously~\cite{whitepaper, intel2019},
important information about the replacement and update policies have not been shared publicly.
In this work, we share the results of the reverse engineering process and present them in Section~\ref{sec:rvp}.

There are also other important parameters %
used during the operation of hardware prefetchers. For example, cross-page prefetching will be restrained to avoid page faults and the consequent delay. The page checking issue will be discussed in detail in Section~\ref{sec:rvp}. %

\begin{table}[tb]
\footnotesize
\centering
\scalebox{0.9}{
\begin{tabular}{|c|c|c|}
\hline
Intel Prefetcher                                                           & Location & Pattern                                                                          \\ \hline
\begin{tabular}[c]{@{}c@{}}Data Cache Unit\\ (DCU)\end{tabular}            & L1-D     & Next cache line                                                                  \\ \hline
\begin{tabular}[c]{@{}c@{}}Instruction pointer\\ (IP)-stride\end{tabular} & L1-D      & \begin{tabular}[c]{@{}c@{}}Load instructions with \\ regular stride\end{tabular} \\ \hline
\begin{tabular}[c]{@{}c@{}}Data prefetch logic\\ (DPL)\end{tabular}        & L2       & \begin{tabular}[c]{@{}c@{}}128-bytes-aligned \\ pair cache line\end{tabular}   \\ \hline
Streamer                                                                   & L2       &  
\begin{tabular}[c]{@{}c@{}}Several cache lines backward \\ or forward\end{tabular}                                                       \\ \hline
\end{tabular}}
\caption{Documented Intel Hardware Prefetchers. %
}
\label{tb:prefs}
\end{table}

\subsection{Cache timing side-channel attacks}
\label{sec:pp}
Whenever the time taken for the processor to perform certain operations is dependent on secret values, timing side-channels can exist~\cite{Wu2018Timing}. Instruction-based timing side-channels rely on the correlation between the secret and the number of CPU cycles needed to execute an instruction segment. Cache-based timing side-channels exploit the latency gap between the cache and memory subsystems. When the secret value is related to the memory access behavior of the system, attackers have the potential to extract the secret by observing timing differences.

Flush+Reload~\cite{yarom2014flush+} is one of the cache timing side-channels that takes advantage of shared memory between different processes. The attacker first \textbf{flushes} out the shared memory from cache into DRAM. After the victim performs its normal execution, the attacker then \textbf{reloads} the shared memory and observes the timing differences, where addresses that hit in the cache indicate an access by the victim.

The Prime+Probe~\cite{osvik2006cache,percival2005cache} cache side-channel does not require the existence of shared memory. The attacker \textbf{primes} the cache sets with its own data, and \textbf{probes} whether these cache sets are still occupied after the victim program has been scheduled. Prime+Probe is more general than Flush+Reload but it is less noise-resilient, since any activity in the system can evict the priming data as well. Therefore, under circumstances without interference from a significant number of memory accesses, %
the attacker can determine the secret through the use of an inclusive LLC-based Prime+Probe.

Priming the complete LLC cache
might not be easy to achieve by 
accessing a chunk of data whose size is larger than the LLC. This is because most recent microarchitectures divide the LLC into smaller \textbf{slices} to reduce contention and the slice-selection algorithm is not publicly known. The \textbf{eviction set}, as a formal term of priming data, is a collection of addresses that map to the same cache set and slice that guarantees its complete eviction~\cite{systematic15euro, song2019dynamically, theory19sp}. A minimal eviction set (MES) has the number of elements equal to the cache associativity. For example, if the LLC of a processor has 4 slices (normally corresponds to the number of cores) and its associativity is 16, each MES has 16 elements to cover a cache set on a specific slice.

The generic model for constructing the MES~\cite{song2019dynamically} systematically tests among a large candidate address pool to see whether an element can be evicted by them and iteratively shrink its size, which is an extremely %
time-consuming process. However, the LLC is indexed by a part of the bits of physical addresses, which means that the computed MESs can be used as a template and can be reused directly. Other methods aim to exploit the CBox hardware counter per slice stored in Model Specific Registers (MSRs)
to reverse-engineer the slice-selection function~\cite{maurice2015reverse}. This method may not be effective on newer generation of Intel processors due to the CBox modifications. %

In the priming phase, the attacker occupies the LLC cache with the MESs and records their access time, which 
should result in
all cache hit latencies. %
In the probing phase, MESs are accessed again and their new access time is compared with the baseline. If the time variance is larger than a threshold, this MES has elements that have been evicted and it is highly possible that the victim has loaded data into this cache set and slice.

\section{Understanding Intel's IP-stride Prefetcher}
\label{sec:rvp}
The IP-stride prefetcher, as implemented in Intel's Sandy Bridge microarchitecture, was previously described in an Intel white paper~\cite{whitepaper}. Haswell, a newer generation microarchitecture, uses enhanced data prefetchers~\cite{intel2019}, but
the details of these updates remain undocumented.
Prior works~\cite{host, ccs, papp} have attempted to reverse-engineer a number of the characteristics of the Intel IP-stride prefetcher. %
However, in this work, we take an additional step to reverse-engineer the major components of the IP-stride prefetcher in the Haswell and Coffee Lake microarchitectures. %
To the best of our knowledge, this is the first work to reveal the index, update and trigger mechanisms in detail. We additionally investigate the effects of cross-page address prefetching, determine the number of entries in the history table, and reverse engineer the IP-stride prefetcher's replacement policy.

\begin{listing}[tb!]
\begin{minted}{C}
void idx_detect_train(int stride, int train)
{
     for(int i = 0; i < train; i++)
     {
IP_1:     int temp0 = array[i * stride];
     }
     // Not shown: add IP offset using NOPs
IP_2:int temp1 = array[r];
}

\end{minted}
\caption{Microbenchmark pseudo-code for detecting the indexing mechanism of the IP-stride prefetcher. %
}
\label{listing:bench_index}
\end{listing}

\begin{figure}[t]
    \centering
    \includegraphics[width=1\linewidth]{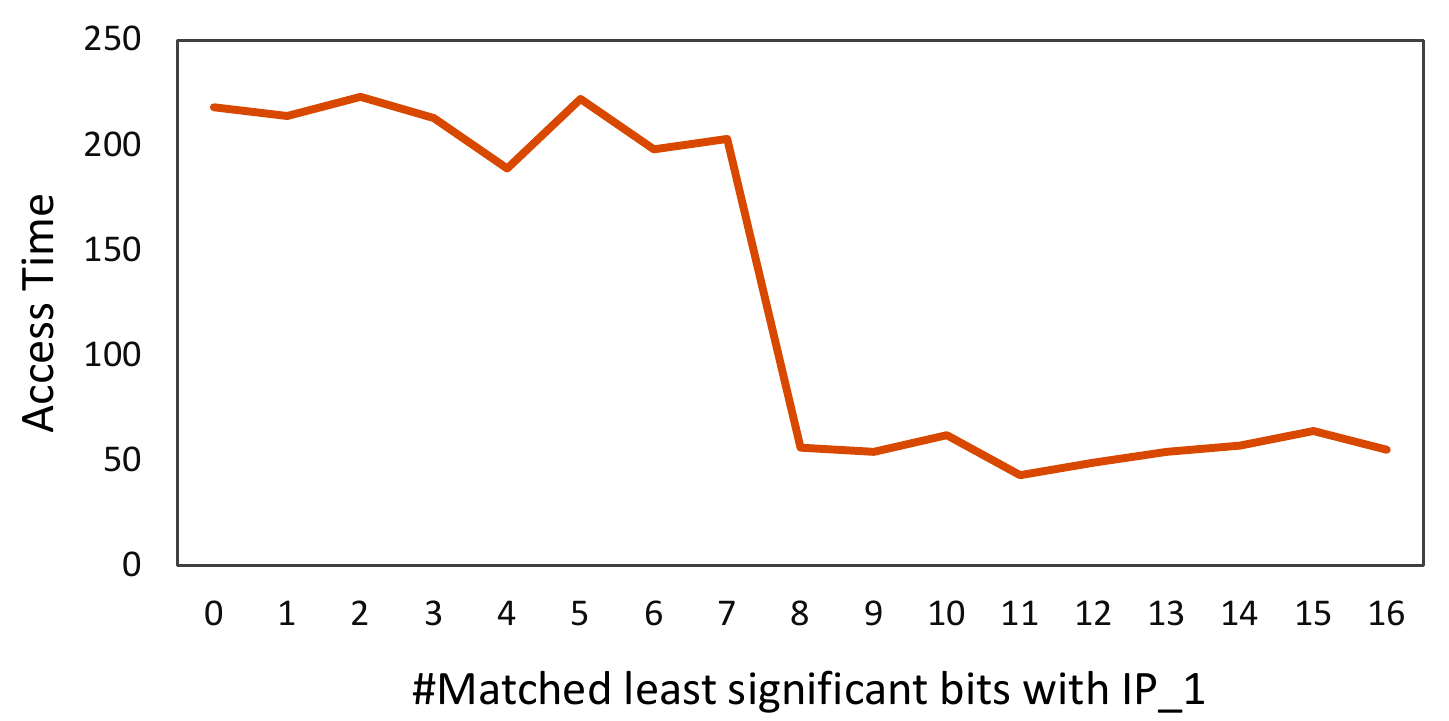}
    \caption{The Coffee Lake IP-stride prefetcher triggering a result on \texttt{IP\_2} when trained with \texttt{IP\_1}. Note that an \textit{Access Time} higher than 120 cycles means that the prefetcher has not been triggered to prefetch the address into cache.%
    }
    \label{fig:bench_index}
\end{figure}

\subsection{Indexing into the IP-stride Prefetcher}
To understand how the IP-stride prefetcher is indexed in hardware, we start by investigating 
the details of the prefetcher's history table.
Since older generations of Intel processors are indexed with the least significant 8-bits of the load instruction address, we first aim to verify whether this is still true in newer processor generations. %
Moreover, we would like to investigate whether there are any other factors that should be taken into account during indexing.

We use a microbenchmark, similar to that shown in Listing~\ref{listing:bench_index}, which first trains the IP-stride prefetcher using \texttt{IP\_1} with a constant multiple cache line-sized stride and then accesses the \textit{r}-th cache line in the \textit{array} with \texttt{IP\_2} (e.g., \textit{mov array[r], rax}). \texttt{IP\_2} is also offset %
such that the least significant \texttt{n}-bits match those of \texttt{IP\_1}. If the load at \texttt{IP\_2} can activate the prefetcher to prefetch \textit{array[r + stride]}, i.e. map to the same entry with \texttt{IP\_1}, %
we conclude that the indexing of the IP-stride prefetcher is dependent on these \texttt{n}-bits of IP.
Figure~\ref{fig:bench_index} shows that the \texttt{IP\_2} \texttt{load} can trigger the prefetcher if its lowest 8-bits are the same as that of \texttt{IP\_1}, confirming our understanding that the IP-stride prefetcher uses the least significant 8-bits to index the entry. Furthermore, this example verifies that the IP-stride prefetcher lacks a tag field to verify the full IP.

\begin{listing}[tb!]
\footnotesize
\begin{minted}[xleftmargin=20pt, linenos]{C}
void const_ip_load(int index, void* array)
{
IP_1:     int temp = array[index];
}

void policy_cs(int st_1, int st_2, int tr_1, int tr_2)
{
    char* array = mmap(4096, ...);
    int i, offset;
    for(i = 0; i < tr_1; i++)
        const_ip_load(i * st_1, array);
    flush(array);
    for(i = 0; i < tr_2; i++)
        const_ip_load(offset + i * st_2, array);
    // test whether the stride now is updated to st_2
    time(array[i * st_2]); 
    // test whether the stride now is still st_1
    time(array[offset + (i -1 ) * st_2 + st_1]); 
}

\end{minted}
\caption{Microbenchmark pseudo-code for detecting the confidence and stride updating policy of the IP-stride prefetcher. 
}
\label{listing:bench_cs}
\end{listing}

\subsection{Confidence and Stride Details}
In the IP-stride prefetcher implementation, the \texttt{confidence} determines whether a prefetch should be triggered and the \texttt{stride} determines the offset of the prefetch. Investigating how these two fields are updated or maintained are crucial in understanding how the IP-stride prefetcher operates.
With this knowledge, it can then be possible develop the \name{}  side-channel. %
To reverse-engineer these details, we design our experiments by training the prefetcher in two phases with different strides, and observing its behavior at each step. 

According to our basic test, the \texttt{confidence} has two bits and the threshold is 2 (we use the \texttt{for} loop shown in Listing~\ref{listing:bench_index}, and set different values for the \texttt{train}. %
After training, we check \textit{array[i*stride]} to see if it is cached). The \texttt{stride} has (1+12) bits, with the most significant bit is used to differentiate between negative and positive strides, while the other twelve bits reflect the maximum stride, which cannot be more than 2KiB ($1 << 12$). It should be noted that the stride of Intel's IP-stride prefetcher does not need to align to a cache line~\cite{intel2019}. %
As a result, the prefetcher requires up to 13 bits to deliver the stride. However, because we train the prefetcher using cache-line-sized data offsets in this work,
a stride of %
7 %
means that
the stride recorded in the IP-stride prefetcher has a length of $7\times64$ bytes, or 7 cache lines in total.

\begin{figure}
     \centering
     \subfloat[][Two training phases with a random offset in between.]{\scalebox{0.3}{\includegraphics{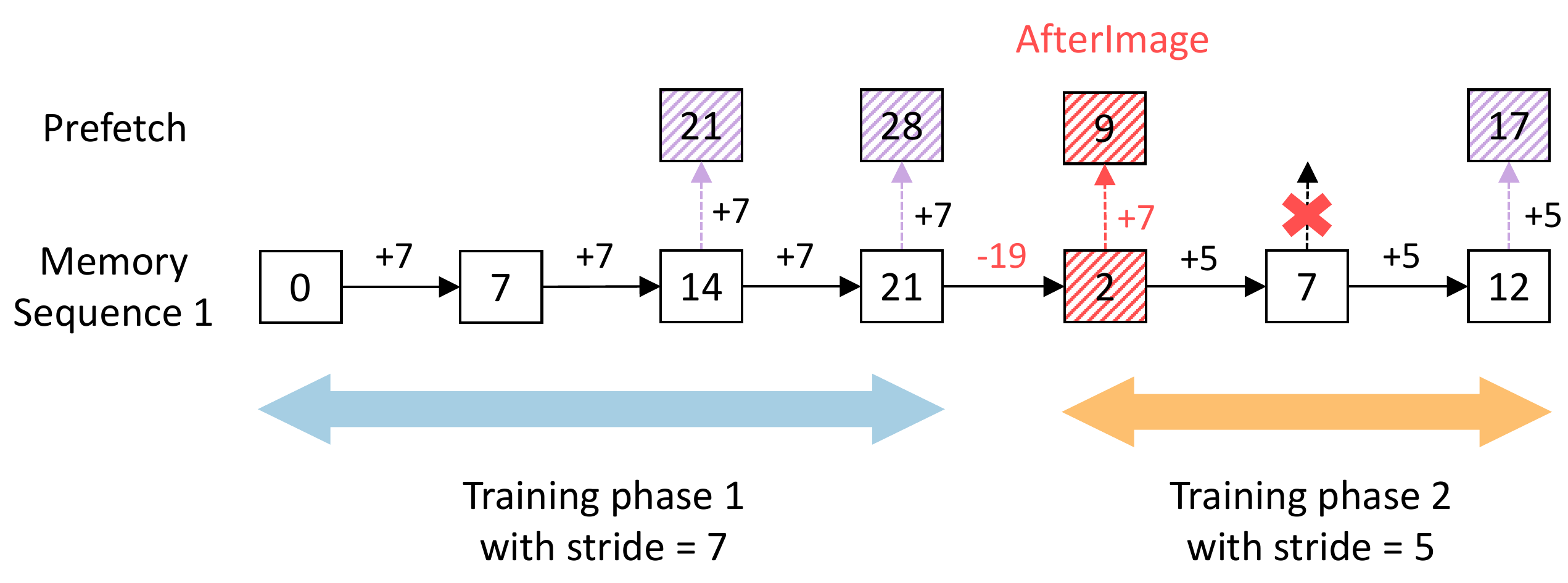}\label{fig:cs1}}}
     
     \subfloat[][The second training phase starts immediately after the first.]{\scalebox{0.3}{\includegraphics{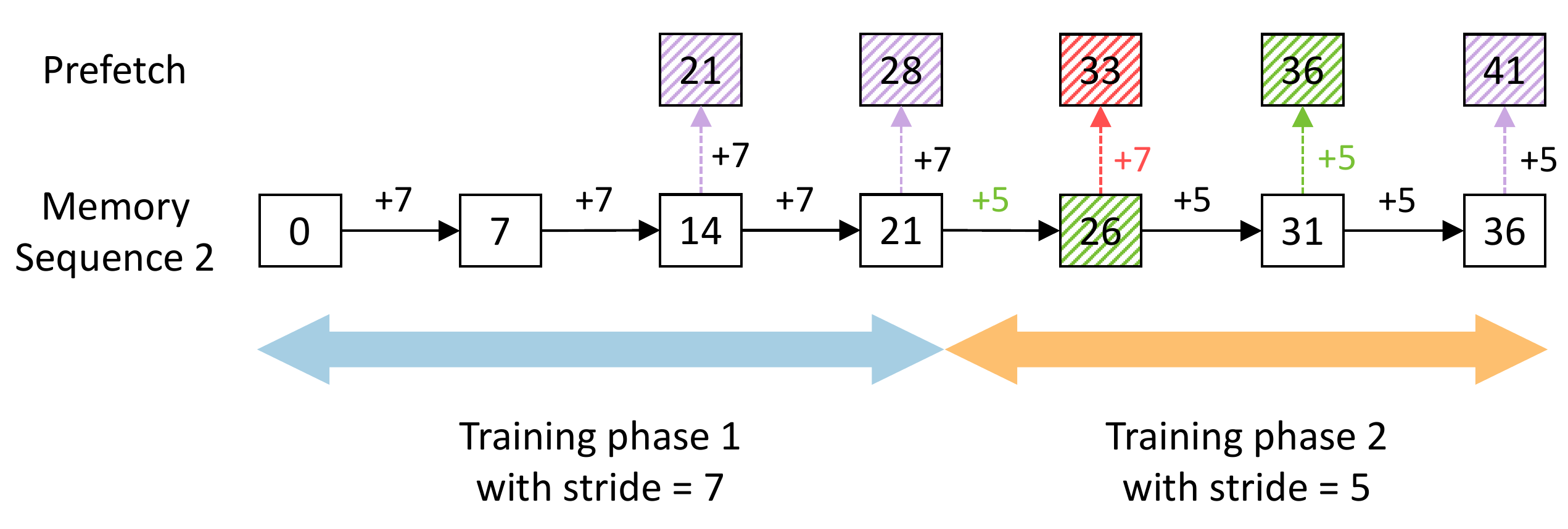}\label{fig:cs2}}}
    
    \caption{Experimental results of the IP-stride prefetcher triggering mechanism on Coffee Lake. Even though the prefetcher sees a new stride value, it will continue to prefetch using the most recently trained stride, enabling \name{}.}
    \label{fig:cs}
\end{figure}

After determining the prefetcher's supported %
\texttt{confidence} and \texttt{stride} values, we use a %
microbenchmark %
(Listing~\ref{listing:bench_cs}) to reverse-engineer the confidence and stride update policy. The microbenchmark first trains the \texttt{IP\_1} \textit{tr\_1} times (\textit{tr\_1} > 2 to guarantee the confidence is equal or larger than the threshold) with stride \textit{st\_1}, and then uses a new stride (\textit{st\_2}) to train the prefetcher \textit{tr\_2} times. Finally, the results from the microbenchmark run allows us to determine which stride is currently being used in the prefetcher, with the %
results listed in Figure~\ref{fig:cs}.

In our experiment, \textit{st\_1} and \textit{st\_2} are set as 7 and 5, respectively. Normally, a stride of \textit{st\_1} will be triggered even after the first iteration of the second loop. This means that
\textit{regardless of whether the new stride is identical to the recorded stride, if the confidence reaches the required threshold, the prefetcher always issues a new prefetch request.} We refer to this behavior as the \textbf{key} component of \name{}. This allows for the triggering of the prefetcher to occur unconditionally, allowing the result to appear in a separate execution context.

For the second iteration of the second loop, no matter how large the value of \textit{tr\_1} is, neither \textit{st\_1} nor \textit{st\_2} is triggered. When we get to the third iteration, the \textit{st\_2} is finally active. However, we discover that if we set the \texttt{offset} directly to the stride of the second phase,
i.e. start the second training earlier, the prefetcher will become fully trained %
at the second iteration. These results imply that the \texttt{stride} will always be updated as $current\ address\ -\ last\ address$, and once the newly computed \texttt{stride} is different from the previous stride, the \texttt{confidence} will be reset to 1 at the same time. Therefore, for a \texttt{load} instruction with an IP and a request to $current\ address$, the workflow inside the prefetcher is shown in Algorithm~\ref{algo::policy}.

The complete update policy for the \texttt{confidence} and \texttt{stride} can be described as follows. After a new entry is made, the \texttt{stride} and \texttt{confidence} are set to the learning value and 1, respectively, at the second cache miss. If the stride remains constant in the third iteration, the confidence adds 1 (now is 2) and triggers the prefetcher to prefetch the $current\ address\ +\ stride$. If the confidence is larger than 1, the prefetcher will prefetch $current\ address\ +\ stride$, even it finds the $current\ address\ -\ last\ address\ \neq\ stride$ in the next iteration. In the meantime, due to the $current\ address\ -\ last\ address\ \neq\ stride$, the prefetcher updates the stride to $current\ address\ -\ last\ address$, and then reset the confidence to 1 (after prefetching).

\begin{algorithm}[h!]
\SetAlgoLined
\KwData{IP, current\_address}
\eIf{IP tag existed in the history table}
    {
      distance = $current\ address - last\ address$\
      
      \eIf{confidence $\geq$ 2}
      {
        \textbf{Prefetch $current\ address + stride$}\
        
        \eIf{distance != stride}
        {
            stride = distance\
            confidence = 1\
        }{
            \If{confidence != 3}{
                confidence += 1\
            }
        }
      }{
        \eIf{distance != stride}
        {
            stride = distance\
            
            confidence = 1\
        }{
            confidence += 1\
            
            \If{confidence == 2}{
                \textbf{Prefetch $current\ address + stride$}\
            }
        }
      }  
    }{
        $Create\_New\_Entry$(IP, confidence = 0, stride = 0)\
    }
\caption{Confidence and stride updating policy and triggering strategy of the IP-stride prefetcher.}
\label{algo::policy}
\end{algorithm}

\subsection{Page Boundary Checking}
Modern processors often use virtual memory management with memory paging support. According to previous studies~\cite{host, intel2019}, when an instruction loads data ($current\ address$) from a new \texttt{page}, the prefetcher invalidates the entry and will need to re-learn the stride. %
To understand how cross-page accesses affect this prefetcher in newer generations, and to allow %
\name{} to be applied to a broad set of scenarios across different execution contexts, %
we next broaden our investigation to reverse-engineer the undocumented elements of the page checking policy.

The benchmark shown in Listing~\ref{listing:bench_page} first builds two memory pools, named \texttt{recl\_array} and \texttt{lock\_array}. \texttt{recl\_array} is allocated without \texttt{MAP\_LOCKED}, which is a resource-saving pool that automatically reclaim used physical page frames. The \texttt{lock\_array}, on the other hand, will always lock the page frame. We leverage \texttt{IP\_1} and \texttt{IP\_2} to train the prefetcher with given stride on one page (e.g., \textit{p}-th page), and then access the next \textit{offset}-th page, and verify whether the target address (i.e., \textit{recl\_/lock\_array[p + offset + stride])} is in the cache or not. 

Table~\ref{tb:page} shows the result of this experiment. 
The first column is the offset between the testing page and the training page. The second and third column indicates whether these testing pages have the same physical or virtual address with the training page or not. The last column represents the testing results, i.e. successfully prefetched or not.
Despite the fact that the destination address in \texttt{recl\_array} spans several logical page boundaries, the prefetcher does not invalidate the entry \texttt{IP\_1} and they are all successfully triggered. If the physical page frame is crossed, the prefetcher can invalidate the entry and re-learn the stride and confidence. More specifically, if the newly accessed page is the next page of the learning page (\textit{(p+1)}-th page) and this page misses in the TLB, the first access for this page will create the page table entry and will not impact the prefetcher status (e.g., decrease the confidence). The second memory access on the \textit{(p+1)}-th page then can directly activate the prefetcher to prefetch $current\ address\ +\ stride$. If the page mapping hits in TLB, the prefetcher will be activated immediately. However, for pages further into the future, i.e. \textit{offset > 1},
prefetching will be prohibited.
We infer that the next page is special because of the use of the next-page prefetcher that was introduced in the Haswell microarchitecture~\cite{npp}.

As a result, the prefetcher uses the page frame to determine whether the new address crosses the page boundary and processes the next page separately.

\begin{listing}[tb!]
\footnotesize
\begin{minted}[xleftmargin=20pt, linenos]{C}
void two_ip_loads(int index, void* array1, void* array2)
{
IP_1:     int temp0 = array1[index];
IP_2:     int temp1 = array2[index];
}


void page_policy(int offset, int stride)
{
    char* recl_array = mmap(n * 4096, ...);
    char* lock_array = mmap(n * 4096, MAP_LOCKED, ...);
    // do not cross page
    for(int i = 0; i < 4; i++)
        two_ip_loads(i * stride, recl_array, lock_array);
    
    two_ip_loads(offset, recl_array, lock_array);
    time(recl_array[offset + stride]);
    time(lock_array[offset + stride]);
}

\end{minted}
\caption{Microbenchmark pseudo-code for detecting the page checking strategy of the IP-stride prefetcher.}
\label{listing:bench_page}
\end{listing}

\begin{table}[]
\centering
\scalebox{0.65}{
\begin{tabular}{|c|c|c|c|c|c|c|}
\hline
\multirow{2}{*}{Offset} & \multicolumn{2}{c|}{Physical Addr} & \multicolumn{2}{c|}{Virtual Addr} & \multicolumn{2}{c|}{Prefetch Target Addr} \\ \cline{2-7} 
                              & recl\_array        & lock\_array       & recl\_array       & lock\_array       & recl\_array       & lock\_array      \\ \hline
1 Page                            & same             & different       & different       & different       & \cmark             & \cmark            \\ \hline
2 Pages                            & same             & different       & different       & different       & \cmark             & \xmark           \\ \hline
3 Pages                            & same             & different       & different       & different       & \cmark             & \xmark           \\ \hline
4 Pages                            & same             & different       & different       & different       & \cmark             & \xmark           \\ \hline
\end{tabular}}
\caption{The Coffee Lake IP-stride prefetcher triggering results on different logic pages and physical page frames.}
\label{tb:page}
\end{table}

\begin{listing}[tb!]
\footnotesize
\begin{minted}[xleftmargin=20pt, linenos]{C}
void n_ip_loads(int index, int N, void* array)
{
IP_1:     int temp1 = array[4096 + index];
 ...               ...
IP_N:     int tempN = array[4096*N + index];
}

void num_entry(int N, int stride, int offset)
{
    //contains N pages
    char* array = mmap(N * 4096, MAP_LOCKED, ...);
    for(int i = 0; i < 5; i++)
        n_ip_loads(i * stride, N, array);
    
    n_ip_loads(offset, N, array);
    for(int i = 0; i < N; i++)
        time(array[4096*i + offset + stride]);
}
\end{minted}
\caption{Microbenchmark pseudo-code for determining number of entries of the IP-stride prefetcher.}
\label{listing:bench_size}
\end{listing}

\begin{figure}[t]
    \centering
    \includegraphics[width=1\linewidth]{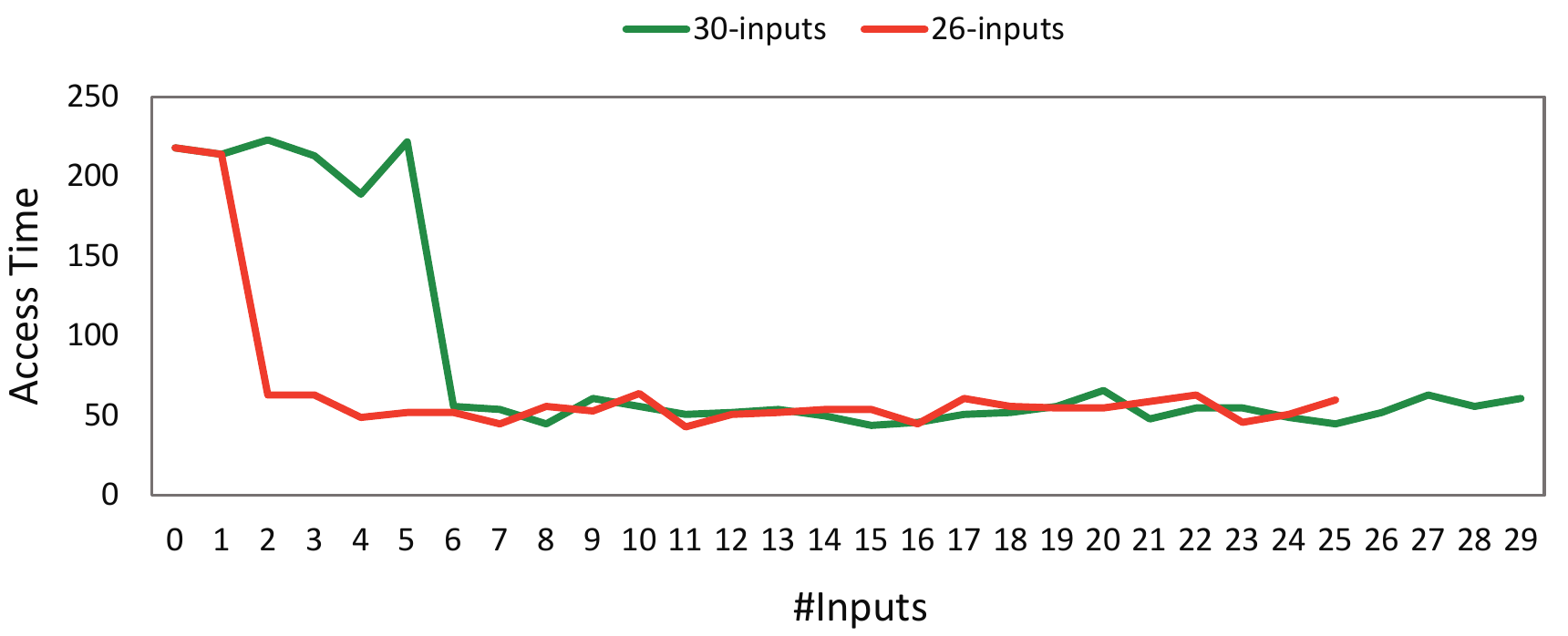}
    \caption{The IP-stride prefetcher triggering results for 26 IPs and 30 IPs to determine the prefetcher's number of entries on Coffee Lake. %
    }
    \label{fig:bench_size}
\end{figure}

\begin{figure}[t]
    \centering
    \includegraphics[width=1\linewidth]{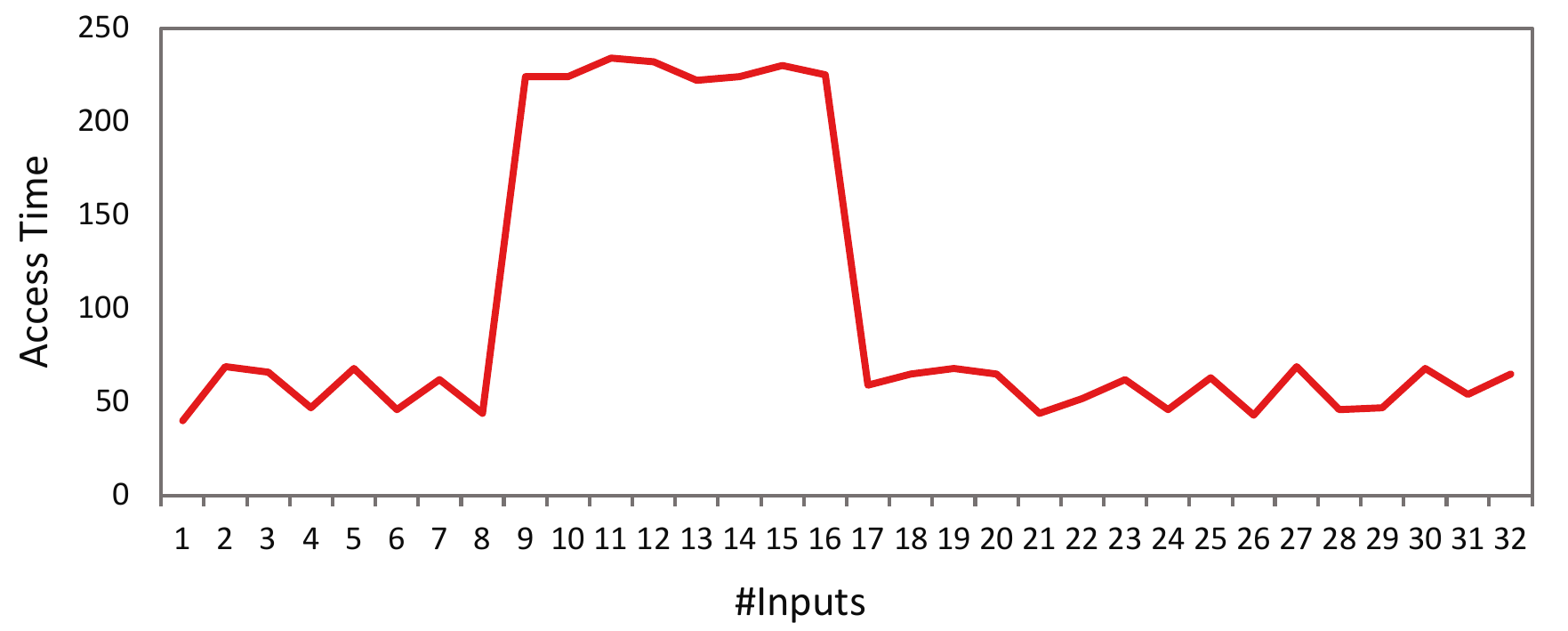}
    \caption{The IP-stride prefetcher triggering result for 32 IPs, with 8-16th IPs revisited, to demonstrate the prefetcher's replacement policy on Coffee Lake.}
    \label{fig:bench_rp}
\end{figure}

\subsection{Number of Prefetch Entries}
\label{sec:num_pf}
The number of entries represents the maximum number of IPs and their corresponding stride that the prefetcher can remember. This will affect how we search for a matched IP when target IP is unknown, as we will show in \ref{sec:v3}, since the hardware capacity is fixed.

We construct a microbenchmark (See
Listing~\ref{listing:bench_size})
that executes a loop with a %
varying number of \texttt{load} instructions. Every \texttt{load}'s least significant 8-bits of its IP are unique, and their data access patterns are constant. After training each of these IPs on different page frames (to avoid false-positives), we re-access them and test the access time to \textit{page\_t[offset + stride]} ($0 \leq t \leq n$) to determine if they can still activate the prefetcher.

The experimental result is shown in Figure~\ref{fig:bench_size}. Specific IPs are no longer able to trigger the prefetcher. More specifically, if the number of test IPs is 26, the first two IPs will no longer be able to %
activate the prefetcher. If the number of test IPs is 30, the first six IPs cannot trigger the prefetcher. As a result of the prefetcher's restricted size, some IPs get evicted. Thus, the number of prefetcher entries is the number of IPs that can still activate the prefetcher after training all of them, which is 24, in our case.

\subsection{Prefetcher Replacement Policy} %
\label{sec:prp}
Since the capacity of IP-stride prefetcher is limited, i.e. 24 as seen in in Section~\ref{sec:num_pf}, when new load instructions are present, some entries should be replaced following a specific replacement policy. Generally, we have found that as long as the load instruction of interest were executed recently, it will remain in the hardware and later trigger a prefetch, which is sufficient to satisfy \name{} in most cases. However, we provide additional information about the IP-stride prefetcher that 
may benefit future researchers.

As we only see the most recent IPs being evicted, to determine whether the prefetcher's replacement policy is First-In-First-Out (FIFO) or least recently used (LRU), we update Listing~\ref{listing:bench_size} by adding a number of \texttt{jmp} instructions. The number of test IPs is increased to 32 and 32 page frames are allocated for the training of each IP. The first 24 IPs will be trained on various page frames to occupy the whole table, and then the caches will be flushed. Next, the first 8 IPs will be re-trained to update them to a more recently used position. Following that, we train another 8 new IPs to evict some entries out and flush the cache once again. Finally, we execute these 32 load instructions again to read a random cache line $L$ in the corresponding page and see if the $(L + stride)$-th cache line is prefetched. The first eight IPs should have been evicted if the prefetcher uses a FIFO policy. If not, these IP addresses should still be able to trigger prefetching. The experimental result is shown in Figure~\ref{fig:bench_rp}. We observe that the evicted IPs are between the 9th and 16th position, indicating that the IP-stride prefetcher is using a form of the LRU replacement strategy. In addition, because the replacements have always been contiguous, 
it follows that it will most likely not use a
tree-based pseudo-LRU (PLRU) replacement policy. %
Further, as a true LRU implementation can be expensive to implement in hardware, we suspect that the hardware is implementing a Bit-PLRU-based replacement policy.

\subsection{IP-stride Prefetcher across Different Generations} %
Using our microbenchmarking methodology, we reverse-engineer the IP-stride prefetcher on Haswell and Coffee Lake and demonstrate the same conclusions. Because the microarchitectures of Sky Lake and Kaby Lake are identical to that of Coffee Lake~\cite{Abel20}, we infer that the Haswell, Sky Lake, Kaby Lake, and Coffee Lake microarchitectures all use the same IP-stride prefetcher, implying that they are all vulnerable to \name{}.

\begin{figure}[t]
    \centering
    \includegraphics[width=1\linewidth]{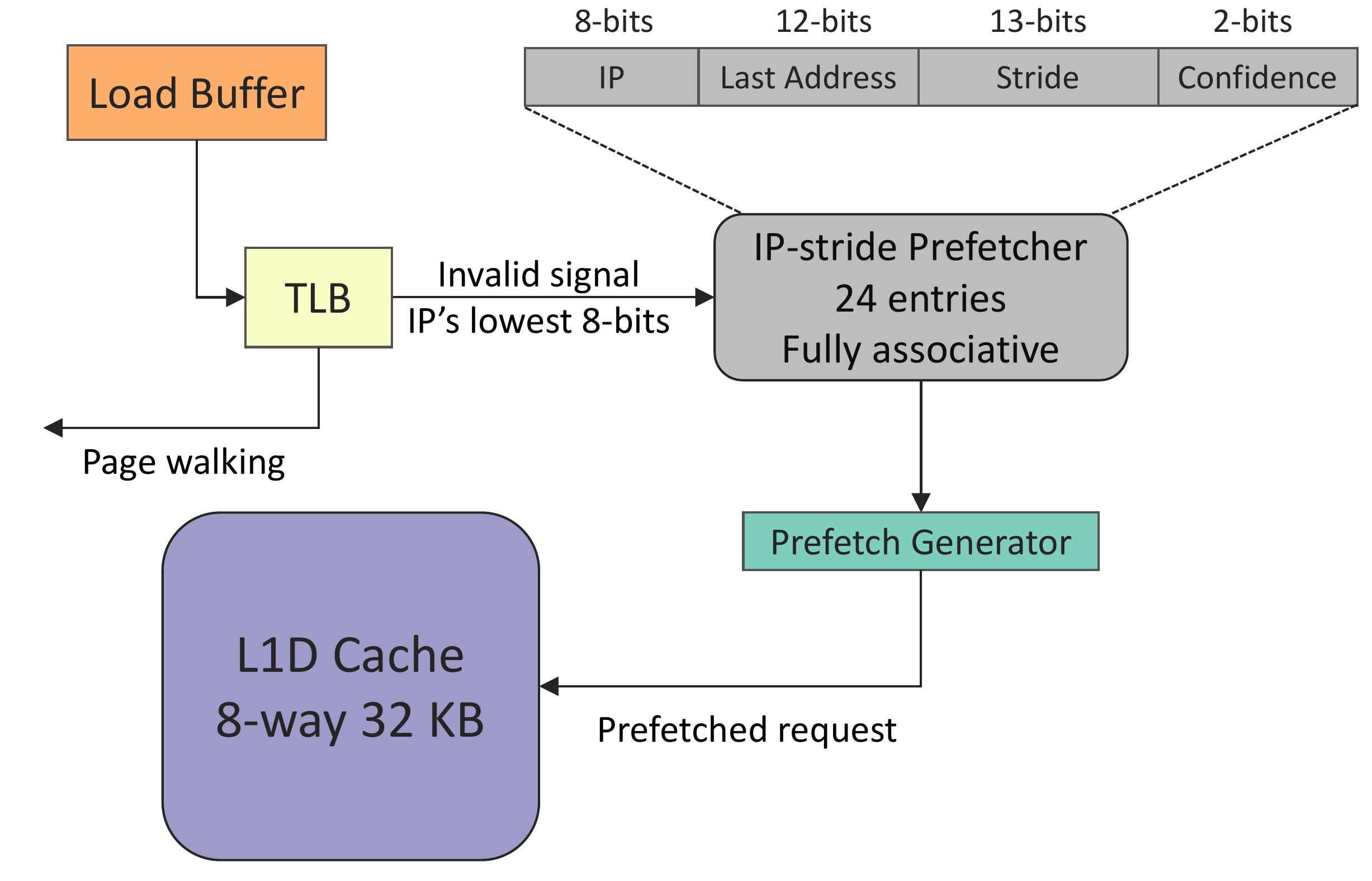}
    \caption{The reverse-engineered microarchitecture of Intel's advanced IP-stride prefetcher found in Haswell, Sky Lake, Kaby Lake, and Coffee Lake. %
    }
    \label{fig:archip}
\end{figure}

Figure~\ref{fig:archip} depicts the microarchitecture of Intel's current IP-stride prefetcher. Compared to the older generation~\cite{whitepaper}, the optimized prefetcher decreases the number of entries from 256 to 24, and the direct-mapped indexing approach is replaced with a fully associative indexing that most likely uses a Bit-PLRU replacement policy. %
The \texttt{last address}, \texttt{stride}, and \texttt{confidence} all have the same structure as before. The page checking policy has also been modified to allow prefetching in the next page frame. Our reverse-engineering efforts have been the first, to the best of our knowledge, to present the page policy, confidence and stride update strategy of the Intel IP-stride prefetcher.

\section{Leaking Secrets with the IP-stride Prefetcher}
\label{sec:attack}

\begin{listing}[tb!]
\begin{minted}[xleftmargin=20pt, linenos]{C}
if(secret)
    int temp0 = array[address]; 
else
    int temp1 = array[address]; 

\end{minted}
\caption{Target minimal vulnerable code region of the victim. %
}
\label{listing:vcode}
\end{listing}

In this section, we present how \name{} leaks control flow information across different regions of isolation %
through the careful training of the
IP-stride prefetcher.
We base \name{} on the characteristics discussed in Section~\ref{sec:rvp} and three key observations with respect to the IP-stride prefetcher's state, to be discussed below, each of which forms the basis for a variant of the side-channel.
We start with
the targeted %
code region example, as shown in %
Listing~\ref{listing:vcode}. %
When the program's control flow is determined by a secret value, and load instructions exist in the if-else branch, \name{} can extract it and leak it. %
In the following sections, we detail the \varinum{} key observations of this work, and introduce the \varinum{} variants of \name{}: (a) attack the victim from same address space but different code regions, (b) attack the victim from different address space, and (c) attack the Operating System (OS) across the user-kernel boundary.
When successfully executed, these techniques render no observable control flow changes or anomalous branch-speculative executions, and therefore have the ability to bypass several recently proposed defenses against transient attacks, e.g., control flow integrity~\cite{koruyeh2020speccfi, loughlin2021dolma}, and access control restrictions~\cite{qi2021spectaint,taram2019context, wang2019oo7}. %

\subsection{\name{} variant 1}
\begin{center}
\fcolorbox{black}{gray!10}{\parbox{.9\linewidth}{\textbf{Observation 1}: The IP-stride prefetcher trained by IP1 can be triggered by IP2, as long as the least significant 8-bits of IP2 match those of IP1. }}
\end{center}

The goal of this variant is to leak the victim's control flow from different code regions in the same address space. %
We first design a \texttt{gadget} (see Listing~\ref{listing:gadget}) that consists of two \texttt{load} instructions with different IPs. The least significant 8-bits of IPs of these two \texttt{load} instructions are tailored to match the memory access instructions in if-path and else-path in the victim's %
code region, respectively, i.e., line 3 and line 8 in Listing~\ref{listing:vcode}.
By executing the \texttt{gadget} with two constant-strided memory address sequences for two \texttt{load} instructions, we push two entries into the IP-stride prefetcher for these two IPs and saturate their confidence counters for their specific strides. Based on the \textbf{Observation 1}, the victim will trigger the IP-stride prefetcher to prefetch data located at \textit{current\ address\ +\ stride}. The key to this method is that even if the base address the victim uses is unknown, \textbf{one can distinguish the control flow, and therefore the secret condition, by the stride value alone}.

\begin{listing}[tb!]
\begin{minted}[xleftmargin=20pt, linenos]{C}
for(int i = 0; i < 3; i ++)
{ 
    IP offset1
    // to match if-path
    int temp0 = array[i * stride1];
    IP offset2
    // to match else-path
    int temp1 = array[i * stride2];
}
\end{minted}
\caption{\texttt{Gadget} used in variant 1 and variant 2. Note that if-else conditions are not required in the \texttt{gadget} used by the attacker, only the instruction addresses of the loads need to match. The use of different strides (\texttt{stride1} and \texttt{stride2}) allow the attacker to differentiate the two cases. %
}
\label{listing:gadget}
\end{listing}

To accomplish the attack, both Prime+Probe~\cite{osvik2006cache} and Flush+Reload~\cite{yarom2014flush+} can be adopted to effectively observe the stride triggered by the victim that is now present in the cache. %
In terms of a Prime+Probe implementation, after mis-training, we prime the LLC sets (with methods introduced in Section~\ref{sec:pp}) and record the access time for each minimal eviction set (MES) to an array called \texttt{prime\_time}. After victim execution, we probe the LLC on a cache-line granularity and record the access time for MESs to another array, named \texttt{probe\_time}. With Eq.~\ref{eq:atv}, the attacker can simply observe the timing variations ($tv$) in the system to collect which cache sets have elements that have been evicted, i.e., which cache sets have been accessed by the victim. %

\begin{equation}
\label{eq:atv}
        tv = prime\_time - probe\_time
\end{equation}

\subsection{\name{} variant 2}
\label{sec:v2}
\begin{center}
\fcolorbox{black}{gray!10}{\parbox{.9\linewidth}{\textbf{Observation 2}: The IP-stride prefetcher is shared by several processes that operate on the same physical core. Process B may utilize the entry of the IP-stride prefetcher trained by process A if their least significant 8-bits of IP match.}}
\end{center}

\name{} variant 2 demonstrates the ability %
to leak the control flow of the victim from a different address space. We implement variant 2 using the Flush+Reload side-channel, where the secret-dependent load instructions of the victim accesses data in shared memory.

\begin{figure}[htbp]
    \centering
    \includegraphics[width=0.85\linewidth]{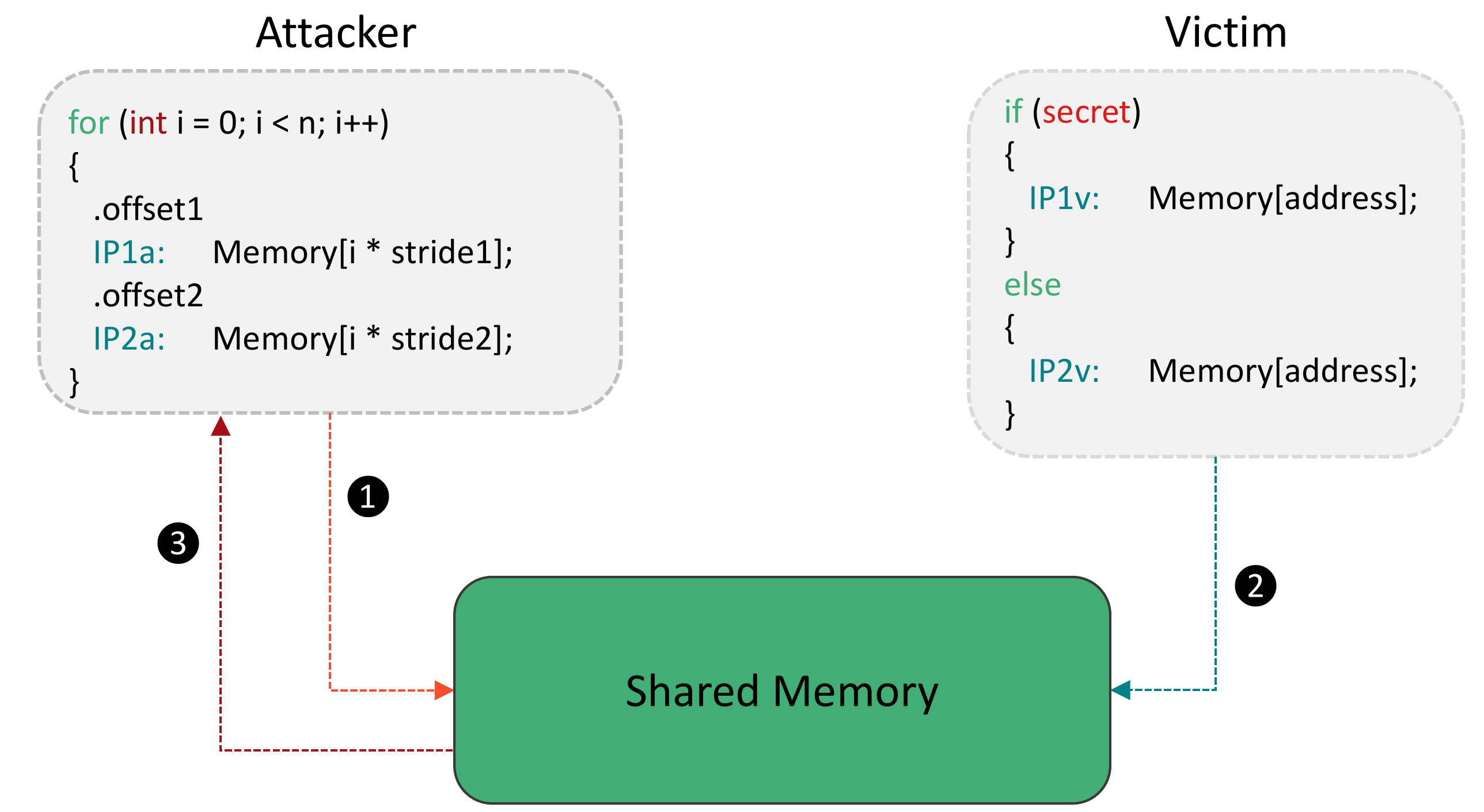}
    \caption{Variant 2: Mis-training the IP-stride prefetcher from another process. Note that code shown in the attacker's region represents our \texttt{gadget} and that $IP1a$ matches $IP1v$ and $IP2a$ matches $IP2v$.}
    \label{fig:v2_flow}
\end{figure}

\begin{figure}[htbp]
    \centering
    \includegraphics[width=0.85\linewidth]{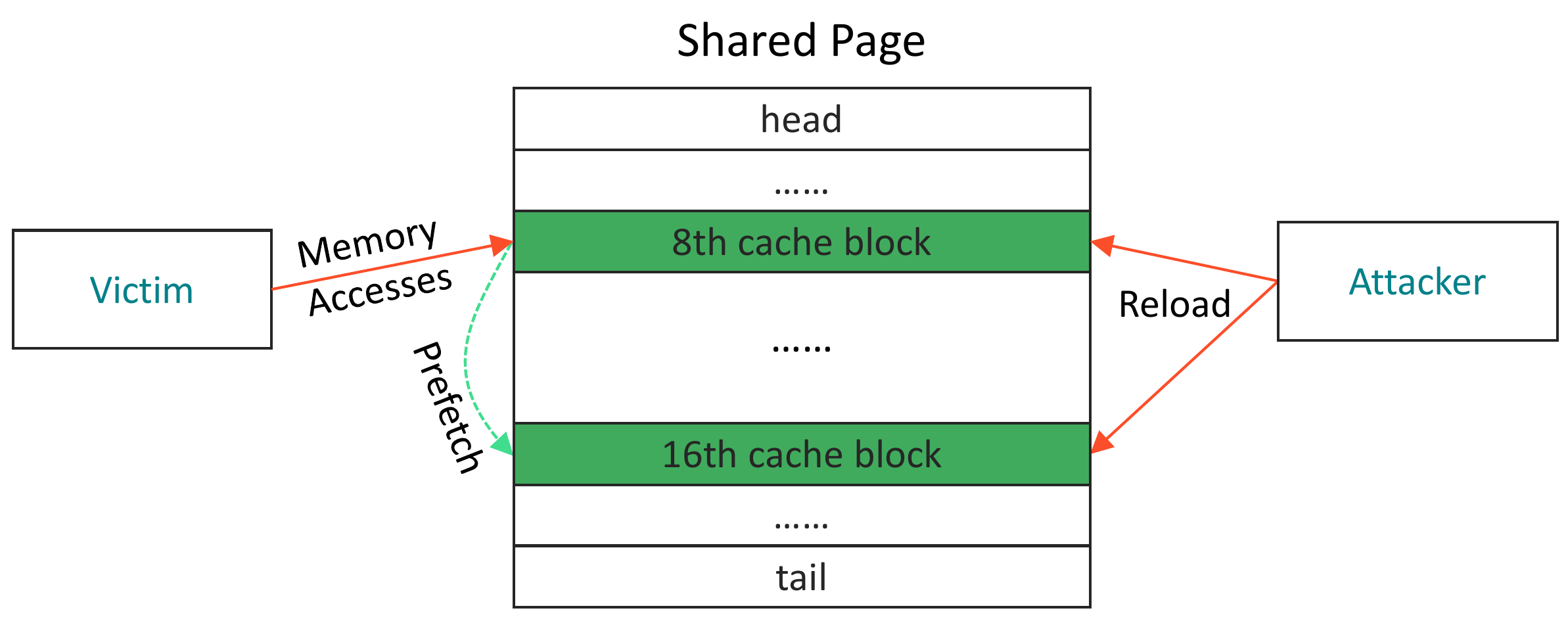}
    \caption{Example of stride detection in a shared memory. The attacker detects that both the 8th and 16th cache line in the shared page are cached, and observes a stride at 8, which matches the stride that is used to train the prefetcher. Note that training of the prefetch stride by the attacker is not shown. %
    }
    \label{fig:v2_stride}
\end{figure}

To attack the victim from another process, we build %
the \texttt{gadget} used in variant 1 in our %
address
space and again train the two %
IPs with distinct stride sequences (e.g., 8 for $IP1a$ and 11 for $IP2a$) (step \circled{1} in Figure~\ref{fig:v2_flow}). %
After training the prefetcher
to a high confidence, the shared pages is then flushed using \texttt{clflush} instruction. %
If the victim process executes the secret-dependent branches (step \circled{2} in Figure~\ref{fig:v2_flow}), based on the \textbf{Observation 2}, as the lowest 8-bits of secret-dependent memory instructions' IPs ($IP1v$ or $IP2v$) will be indexed to an entry that is trained by $IP1a$ or $IP2a$, the IP-stride prefetcher will prefetch the cache line from $current\ address\ +\ stride$ into the cache. After victim executed the interested branch, we can reload the shared memory and check whether the pre-set stride exists among the cached lines. %
Because the $IP1a$ and $IP2a$ are trained with different strides, we readily discover the victim's run-time control flow (see Figure~\ref{fig:v2_stride}).

\subsection{\name{} variant 3}
\label{sec:v3}
\begin{center}
\fcolorbox{black}{gray!10}{\parbox{.9\linewidth}{\textbf{Observation 3}: When a process switches between user and kernel privilege modes, the trained entries of the IP-stride prefetcher is retained. %
}}
\end{center}

The strict isolation between kernel and user mode protects privileged hardware and system status from being exposed to users. %
Sensitive information in the kernel represents the data that should not be visible to an arbitrary user, e.g., tokens, passwords, and encryption keys. This variant demonstrates how the IP-stride prefetcher can bridge the isolation gap between kernel and user space which leads to a side-channel to potentially leak kernel secrets. 

Our third variant is built on top of Flush+Reload. To establish this kernel-user side-channel, we first create a load instruction with the same least significant 8 bits of the IP as found in the target code of the kernel function (in this example, a system call). We train this IP in the user's address space %
with strided memory addresses. After training, we use the \texttt{clflush} instruction to write back %
the shared memory from all levels of cache. %
Next, we trigger a system call and switch into kernel mode.
During the syscall service, if the branch in our target vulnerable code segment is resolved as $taken$, the following \texttt{load} instruction can trigger a prefetch with our designated stride, which completely happens in the kernel mode.
The final step is to reload the shared memory and detect the access time to see whether the trained access, and strided offset, exists in the cache. %

\begin{listing}[tb!]
\begin{minted}[xleftmargin=20pt,linenos]{C}
void IP_matching(void * addr, int group_num)
{ asm (
    "group0:"
    //jump to the specific group
    "cmp 0, group_num" 
    "jne group1"
    //create 24 IPs
    ".rept 24"  
    "mov (addr), rax"
    //train each IP on different page
    "add 0x1000 addr" 
    ".endr"
    "group1:"
    ...
) }
\end{minted}
\caption{IP matching function.} 
\label{listing:v3IPsearching}
\end{listing}

\textbf{IP matching.}
Since IPs of system call functions are normally unknown to the user or hard to determine, we cannot directly deduce the \texttt{offset} in the \texttt{gadget}. In this case, we designed an IP search method to create the correctly matched IP as the training object, which has the same function of the \texttt{gadget}.
This IP should have the same least significant 8-bits with the load instruction of the syscall function, which fortunately shrinks the searching space to 256 possibilities. Due to the hardware capacity limitation, we search for the IP in groups, 24 IPs as a group to fit the size of the IP-stride prefetcher as indicated in Section~\ref{sec:rvp}.

Listing~\ref{listing:v3IPsearching} shows the IP creating and searching function. For each round, we train one single group of IPs by looping this function several times with strided $addr$ values and a fixed $group\_{num}$. 24 IPs in this group will be trained simultaneously with the same stride on different pages. When the correct group get trained, the syscall can trigger a strided footprint in the cache if it executes the very load instruction. This process can be repeated for multiple times until a matched group is found in case of too many $not\ taken$ during the testing.

This IP matching method is general when the victim's IP is difficult to be accessed. It completely runs as a normal private function without touching any data to which it is not privileged. Specially for kernel functions, once the the system is booted, the IPs of instructions will not be changed. This searching process among different groups is not necessary if another round of \name{} is launched.

\begin{listing}[tb!]
\footnotesize
\begin{minted}[xleftmargin=20pt,linenos]{C}
void vulnerable_syscall1(void* memory_space)
{
   int num = random();
   if(num)
   {
        char *address = get_address(memory_space);
        Memory[address];
   }
   return 0;
}
\end{minted}
\caption{The customized kernel function.} 
\label{listing:vkc}
\end{listing}

\textbf{Attacking.}
To demonstrate its feasibility, we build a simple kernel function as a straightforward example to demonstrate how the prefetcher can leak information between user and kernel space. Listing~\ref{listing:vkc} presents the customized kernel function. In this function, $num$ represents the secret in the kernel and determines the $if$ branch, in which a \texttt{load} instruction is followed. The syscall function shares memory with user via a $memory\_space$ parameter that allows Flush+Reload to take place. 

After the target group of IPs gets well-trained in the aforementioned phase, the attacker calls \texttt{clflush} instruction to flush the shared chunk of data out of cache. Then it calls this syscall and passes the control right to the kernel. If the branch is taken, the following load instructions will be executed on the same shared memory space. The %
IP-stride prefetcher automatically checks its history table and finds a matched entry with a high confidence value. Therefore, it will send prefetch request to the next address, which is $current\ address\ +\ stride$. When the syscall service is finished, process goes back to the user state. The attacker reloads the data to see which addresses are cached. If two addresses with our selected stride are both hit, we can infer that this branch has been $taken$ by the kernel, and vice versa.

\subsection{\name{} Mitigation Strategies}
\label{sec:leak-mitigation}

We consider defending against variants 2 and 3, as they pose a higher threat to the system and applications, which is similar to Spectre~\cite{koruyeh2020speccfi}. We propose two countermeasures that can be implemented during a context switch, similar to the defenses of previous work~\cite{bourgeat2019mi6}. %
One mitigation that can be deployed immediately, is to train each entry of the IP-stride prefetcher to evict all potential entries that could be used as a side-channel. While this step does require microarchitecture knowledge and will add additional delay during context switches, it provides an immediate defense applicable to today's hardware. As an alternative, we also propose a more lightweight %
defense that requires hardware support. More concretely, we propose a privileged \texttt{clear-ip-prefetcher} instruction that can be used on a context switch to invalidate all entries in the IP-stride prefetcher.
See Section~\ref{sec:evaluation} for the new instruction evaluation.

\section{Experimental Setup}
\label{sec:expset}
\textbf{Experimental environment.}
We perform Proof-of-Concept (PoC) side-channel experiments on Haswell and Coffee Lake machines. The architecture details and OS configuration of these two machines is shown in Table~\ref{tb:conf}.  
\begin{table}[tb]
\small
\centering
\begin{tabular}{|c|c|c|}
\hline
                 & i7-4770                    & i7-9700                      \\ \hline
Architecture     & Haswell                    & Coffee Lake                  \\ \hline
CPU cores        & 4                          & 8                            \\ \hline
Last Level Cache & 8MB                        & 12MB                         \\ \hline
Operating System & Ubuntu 18.04               & Ubuntu 18.04                 \\ \hline
ASLR             & Level-2 Enabled            & Level-2 Enabled              \\ \hline
KASLR            & Enabled                    & Enabled                      \\ \hline
DRAM             & DDR4, 2 x 4G               & DDR4, 2 x 8G                \\ \hline
\end{tabular}
\caption{Architecture and system configurations.}
\label{tb:conf}
\end{table}

\textbf{Gadget building.}
We have used several methods to build the \texttt{gadget} to train the IP-stride prefetcher.
In terms of variant 1 and 2, we directly obtain %
the last 8 bits of the IP of the victim's load instructions from disassembly tools, such as \texttt{objdump}. 
For variant 3, we use a more general IP matching method, discussed in detail in Section~\ref{sec:v3}.

In addition, as the Address Space Layout Randomization (ASLR) or Kernel ASLR (KASLR) on Linux has at least a granulatiry on one page (assuming a page-size of 4KiB), these techniques will not change the least significant 12 bits of the IPs. Since the IP-stride prefetcher uses the lowest 8-bits to index its history, ASLR does not impact IP matching or \texttt{gadget} building. %

\textbf{Interference from other prefetchers.}
Except for the IP-stride prefetcher, the other three hardware prefetchers can introduce false positives into the results by loading additional cache lines unexpectedly. Fortunately, 
these prefetchers do not have the address range reach compared to the
IP-stride prefetcher~\cite{papp}. %
The DCU (next-line) prefetcher prefetches only the next cache line. DPL (adjacent) prefetcher prefetches the previous or next cache line. And the streamer prefetches the previous or next several sequential cache lines. Noise introduced by these prefetchers can be easily distinguished when using large multiples of cache lines as the prefetch trigger (see the next paragraph for details). %

\textbf{Choice of stride.}
Due to the presence of the three other hardware prefetchers, we 
use a stride that is
greater than four cache lines. Additionally, the use of uncommon stride values (e.g., a larger prime number) will %
provide additional
noise resilience because they can be easily differentiated. In our experiments, we generally train the prefetcher with stride values of 7, 11 and 13.

\begin{table}[tb]
\small
\centering
\begin{tabular}{|c|c|c|c|}
\hline
          & Boundary      & Flush+Reload          & Prime+Probe           \\ \hline
variant 1 & code region   & \cmark & \cmark \\ \hline
variant 2 & process space & \cmark & - \\ \hline %
variant 3 & user-kernel   & \cmark & - \\ \hline %
\end{tabular}
\caption{All variants' features and current supported measurement models.}
\label{tb:available}
\end{table}

\begin{figure*}
     \centering
     \subfloat[][attack if-path.]{\scalebox{0.174}{\includegraphics{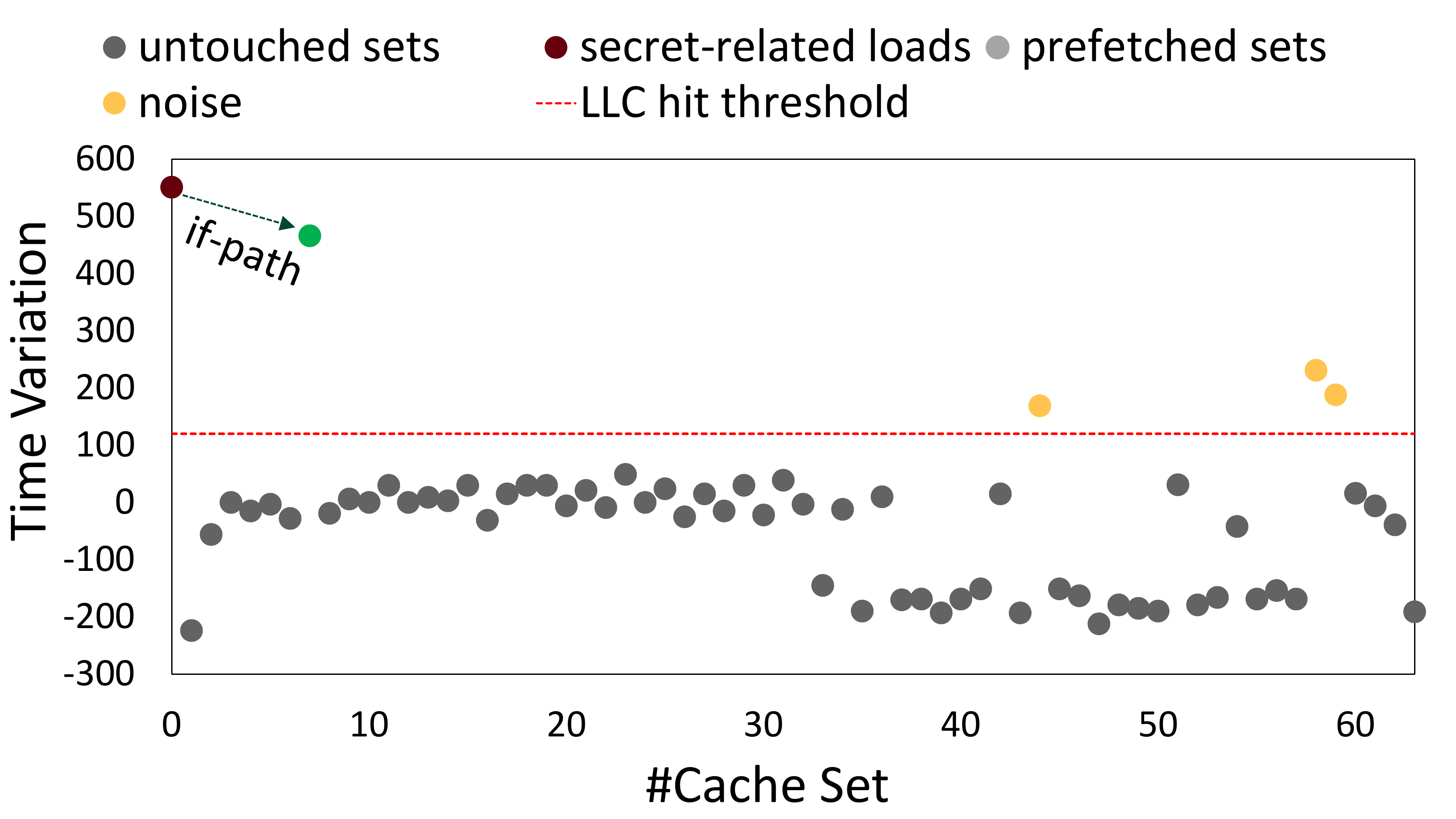}\label{fig:v11}}}
     \subfloat[][attack round-by-round with Prime+Probe.]{\scalebox{0.175}{\includegraphics{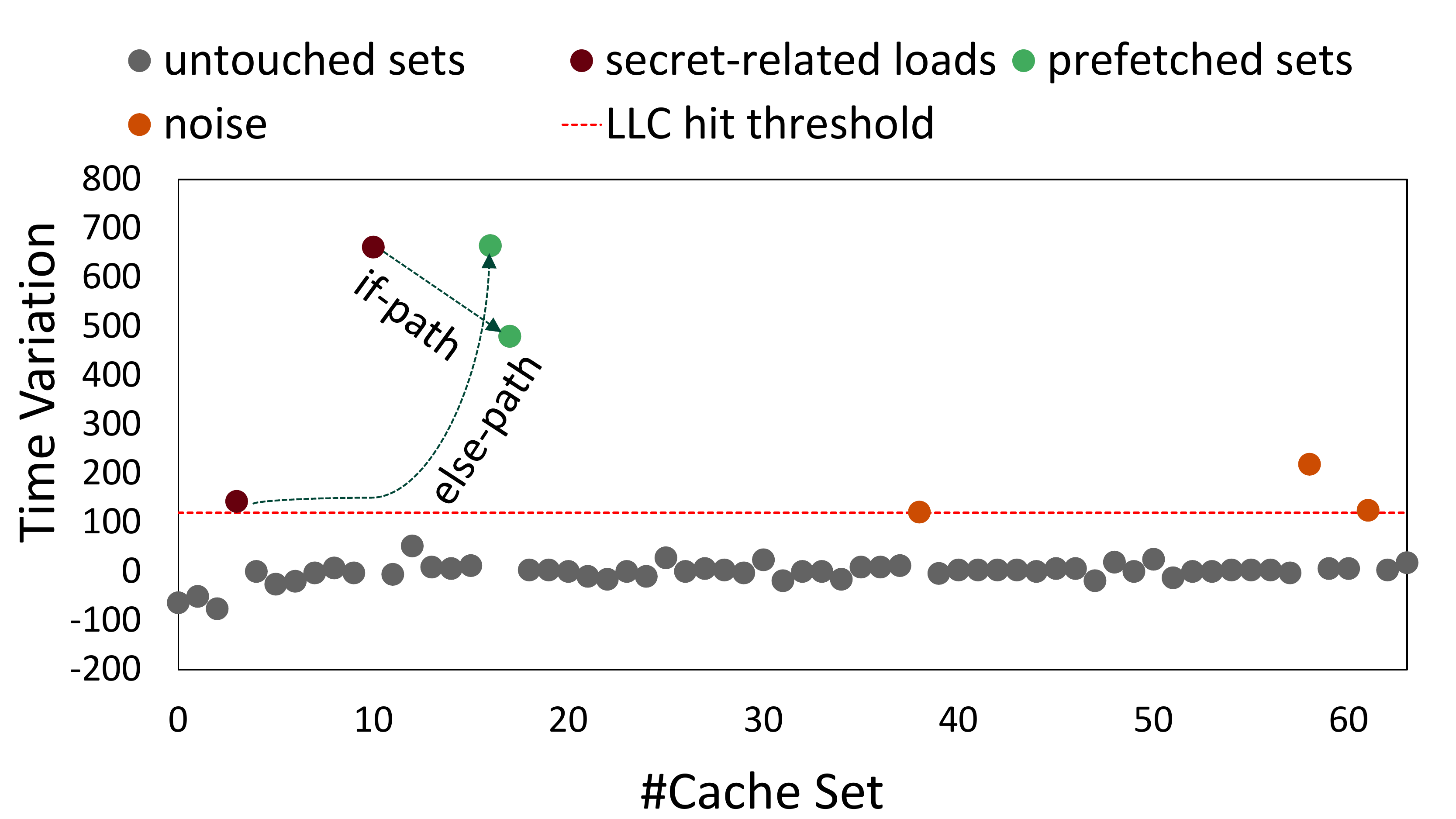}\label{fig:vpp}}}
      \subfloat[][attack round-by-round with Flush+Reload.]{\scalebox{0.172}{\includegraphics{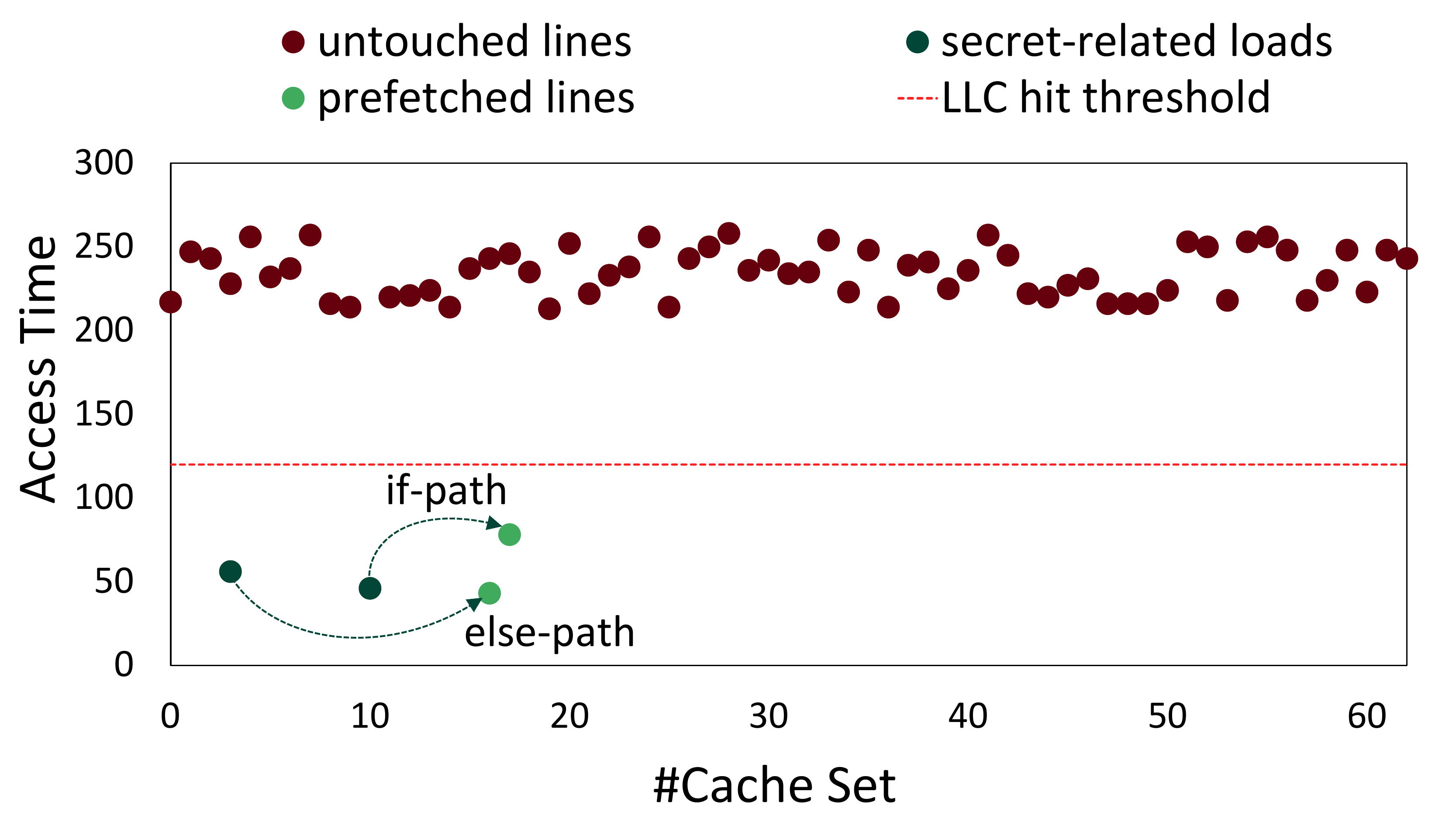}\label{fig:v1fr}}}
     
     \caption{Attack results of \name{} variant 1: (a) single bit extraction from \texttt{if-path} (b) round-by-round extraction from real execution flow with Prime+Probe. (c) round-by-round extraction from real execution flow with Flush+Reload. %
     }
     \label{steady_state}
\end{figure*}

\textbf{Side-channel preparations.}
The Prime+Probe side-channel requires the calculation of minimal eviction sets (MESs) to occupy the LLC cache. We utilize the slice-selection algorithm found in the Haswell microarchitecture~\cite{systematic15euro} to generate the MESs to cover multiple cache sets. If probing is performed with a regular pattern, it will trigger prefetching and introduce many false positives.
Therefore, we organize our MES elements, each with the size of one cache line, into a linked-list data structure. We perform the probing traversal in a pointer-chasing manner, %
and will prevent the
hardware prefetcher from generating requests~\cite{whitepaper,tromer2010efficient}.

For the same reason, Flush+Reload can also introduce noise. %
When we reload a shared page, instead of loading the cache line sequentially, we use the modern version of the Fisher–Yates shuffle algorithm~\cite{shuffle} to randomize the index sequence in the searching range (i.e., [0,63]). %
In addition, we add \texttt{mfence} and \texttt{lfence} when we measure the memory instruction execution time to ensure that measured memory and cache access time is accurate~\cite{host}. 

We note that the insight of this paper, i.e. the IP-stride prefetcher can be intentionally trained to leak information, and its demonstration is our focus. The choice of the underlying timing side-channel used to observe cache activities is not as critical. Except for Prime+Probe and Flush+Reload, other measuring methods such as Flush+Flush~\cite{gruss2016flush+} can apply to this work as well.
As has been introduced in Section~\ref{sec:pp}, Prime+Probe is less noise-resilient than Flush+Reload by nature. We experimentally discover that the process of context switching and user-kernel mode switching in variant 2 and 3 tends to bring intolerable noise to a Prime+Probe side-channel, i.e., over half of MESs are touched by the system. Therefore, we present our experiment results of variant 1 with Prime+Probe and Flush+Reload, and variant 2 and 3 with Flush+Reload in the next section. Table~\ref{tb:available}
presents a list of the successful experiments conducted across the variants of \name{}.

\section{Experimental results}
\label{sec:evaluation}
In this section, we present our experiment results, from variant 1 to 3. For variant 1, we show the results for both \texttt{if} and \texttt{else} branch, on a Prime+Probe and Flush+Reload measurement. For variant 2 and 3, we present Flush+Reload measurement. In the end, we test our success rate for all of the three variants.

Figure~\ref{fig:v11} shows the experimental result of leaking the if-path via Prime+Probe after one round of observation. The 
x-axis represents the
cache set number in our observing page (4KiB page with 64 cache lines), whose distance directly represents the offset between memory addresses in a cache-line-length unit. The 
y-axis shows
the time taken, between the probing phase and priming phase, to access each MES of the cache set.
As depicted in the figure, most cache sets have %
not
been accessed 
as
their 
access latency is lower than 120 cycles. The two cache sets with the highest time delta has a clear stride of 7, demonstrating that the targeted \texttt{load} on the if-path was executed, and triggered an IP-stride prefetch that was trained to be a distance of 7 cache lines.

Note that the prefetcher can prefetch data that is beyond the boundary of an array~\cite{intel2019}, therefore the attacker is still able to observe the stride even the \textit{current\ address} is close to the end of array of the victim (but the prefetched address should not cross page boundary). As the branch predictor typically exhibits a high accuracy in its predictions, we generally see only one stride when the victim executes one of the branches.

We then call the proposed \texttt{gadget} to train the prefetcher for both paths and try to consistently leak control flow round-by-round. %
The synchronization with the victim can rely on simultaneous multithreading or on accurate time-multiplexing. From Figure~\ref{fig:vpp} and Figure~\ref{fig:v1fr}, we observe clear signals (strides) after the victim performed the branch. During the first round, %
we %
see
that the victim took the else-path. The victim then executed the if-path in the following cycle. If the branch is security-related, we then know the secret is \texttt{B'10}. %

From Figure~\ref{steady_state}, we 
observe some noise in the resulting output, but our detection method uses a known stride length, allowing for a high accuracy.%

\begin{figure}[t]
    \centering
    \includegraphics[width=0.85\linewidth]{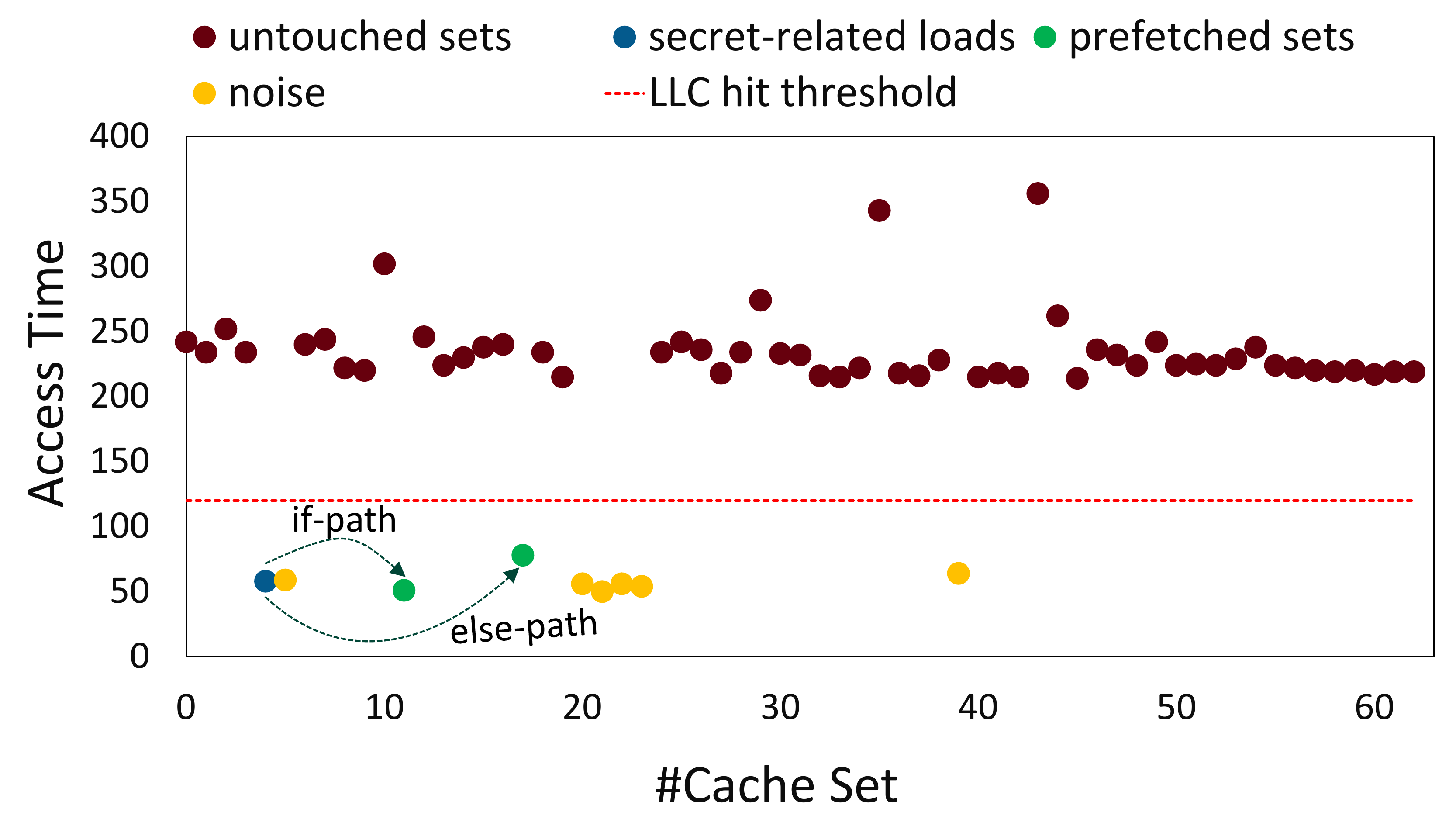}\label{fig:v2at}
    \caption{Attack result of \name{} variant 2.}
    \label{fig:v2re}
\end{figure}

The Proof-of-Concept (PoC) implementation for variant 2 %
demonstrates a cross-process attack. %
In the attacker process, strides for the training of if and else-path are set to 7 and 13, respectively, as in variant 1. %
Figure~\ref{fig:v2re} shows the experiment results extracted by the attacker process, with the cache set again on the x-axis, and the access time for each set on the y-axis.
Using this technique, we are able to 
detect the control flow differences %
by observing the strides for low-access-time cache sets. %

\begin{figure}[t]
    \centering
    \includegraphics[width=0.85\linewidth]{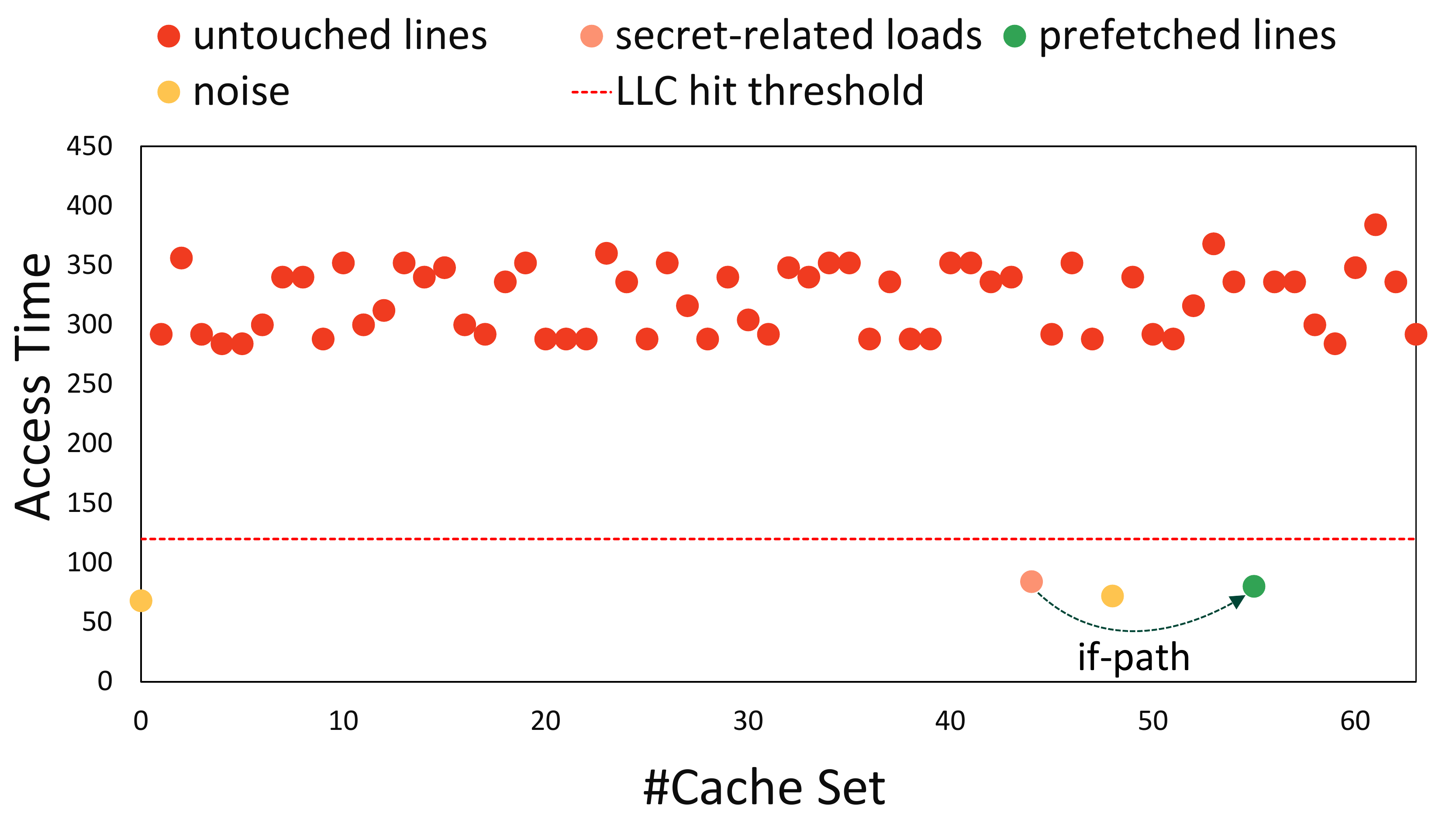}
    \caption{Attack result of \name{} variant 3.}
    \label{fig:v3_fre}
\end{figure}

In terms of \name{} variant 3, we perform the IP search %
as %
described in Section~\ref{sec:v3}, and %
create %
20 groups of 24 load instructions to guarantee the 256 possibilities are covered\footnote{Since there are 480 instructions are created in total, some IPs may fall into more than one group and any of them can serve as the training object.}.
When a matched group is found, we perform the side-channel workflow, as has been introduced in \ref{sec:v3}. %
In this example, we set training stride in the user space to 11 cache lines. In the user reloading phase, the detected stride, shown in Figure~\ref{fig:v3_fre}, indicates that the kernel function executed the if-path and thus the value of \texttt{num} is 1.

To further evaluate the attack success rate, we evaluate %
the %
\varinum{} variants with 200 rounds on a set of sample data. %
We conduct this evaluation on the platform described in Table~\ref{tb:conf}, which is booted in the normal mode without additional user programs running.
The attack success rate of variant 1, variant 2, and variant 3 are 99\%, 97\%, and 91\%, respectively.  %

\textbf{Hardware Mitigation Results.}
To demonstrate the low overhead of our proposed instruction-based defence, we %
implemented it using ChampSim~\cite{champsim}, a cycle-level simulator and the platform used for evaluating a number of computer architecture championships~\cite{dpc3,lin2019branch,pakalapati2020bouquet,peneau2017performance}. %
\begin{figure}[t]
    \centering
    \includegraphics[width=1\linewidth]{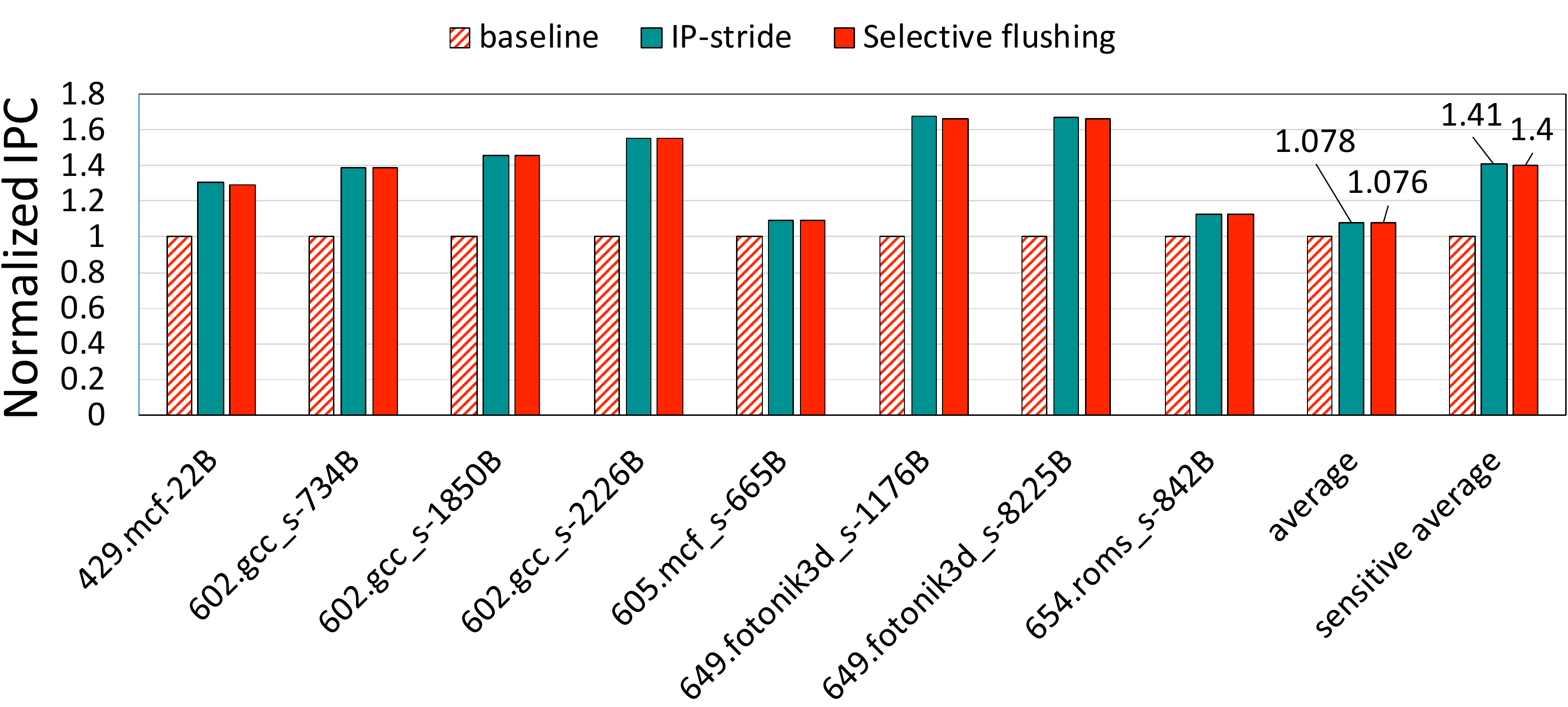}
    \caption{Performance impact for proposed mitigation. Note that the \textit{baseline} configuration represents a processor without prefetching, \textit{IP-stride} denotes the processor with an %
    IP-stride prefetcher, configured as reverse-engineered, and \textit{selective flushing} is the processor that enables our proposed defense with a worst-case period of 10$\pmb{\micro}$s.}
    \label{fig:defense}
\end{figure}

We configure ChampSim to model a Coffee Lake-like processor, and implement the Intel IP-stride prefetcher (based on our reverse-engineering detailed in  Section~\ref{sec:rvp}) to emulate frequent prefetcher flushing due to context switching (every 10 $\micro$s) on different applications in order to emulate a worst-case flushing scenario. %
At each context switch, we flush the IP-stride prefetcher, and %
allow it to 
re-learn the stride. The time to reset the prefetcher is highly dependent by the write ports of prefetcher. For instance, if the we have 2 write ports and 24 entries, which means, every cycle we can overwrite two entries. To clean all slots, we need 12 cycles. To approach the worst case, we set the prefetcher to have only 1 write port, and thus we will need 24 cycles to reset all entries. Using this technique, we then measure the performance 
of the SPEC CPU2006 and 2017~\cite{spec2006,spec2017} benchmarks, with a focus on prefetching-sensitive applications. 
We run 30 billion instructions for each application unless it completes early and then report the normalized IPC.
The overhead introduced by the proposed defense (See Figure~\ref{fig:defense}) is negligible, with an average performance reduction of only 0.7\% for prefetching-sensitive applications (that utilize the IP-stride prefetcher) and 0.2\% across all tested applications. Overhead is low because of the extremely low number of training iterations needed for this prefetcher (which we expect to translate to kernel functions as well).

\section{Discussion}
\label{sec:discuss}
\subsection{Impact from Existing Defenses}
We first discuss the effectiveness of \name{} in the presence of existing defenses that aim to mitigate the threat brought by microarchitectural side-channel attacks. 

\textbf{Control flow integrity.} Many existing defenses are proposed based on control flow integrity~\cite{bhattacharyya2020specrop,koruyeh2020speccfi, loughlin2021dolma}. These protection models will automatically check whether the runtime control flow is deviating from the the control flow graph. If the malicious branch is discovered, those previous proposals will enable their defense (e.g., obfuscated execution, disabling the speculative execution, etc.). However, \name{} does not require speculative control-flow changes; only the backend prefetcher is affected by \name{}, and this occurs during the non-speculative path of execution.

\textbf{Performance counter-based monitoring.} Leveraging performance counters provided by Intel, the defender might be able to identify abnormalities or potential malicious activities in vulnerable hardware components during run-time (e.g., micro-op cache, BPU, L1 cache, LLC, etc.)~\cite{10.1016/j.asoc.2016.09.014,ren2021see,zhang2016cloudradar}. Furthermore, the performance counter provides a performance-efficient way of monitoring the system. However, the sampling frequency and accuracy of performance monitor provided by Intel~\cite{reinders2005vtune} may not be enough to capture the training activities, since \name{} requires just two to three iterations of training at a minimum, and no prefetch request will be generated before it is well-trained. More concretely, Spectre needs around 26,000 cycles to mis-train the BPU~\cite{9007688}, but we need just three cache misses, i.e., 1000 cycles to 2000 cycles (if the first access results in page fault). %

\begin{figure}[t]
    \centering
    \includegraphics[width=0.70\linewidth]{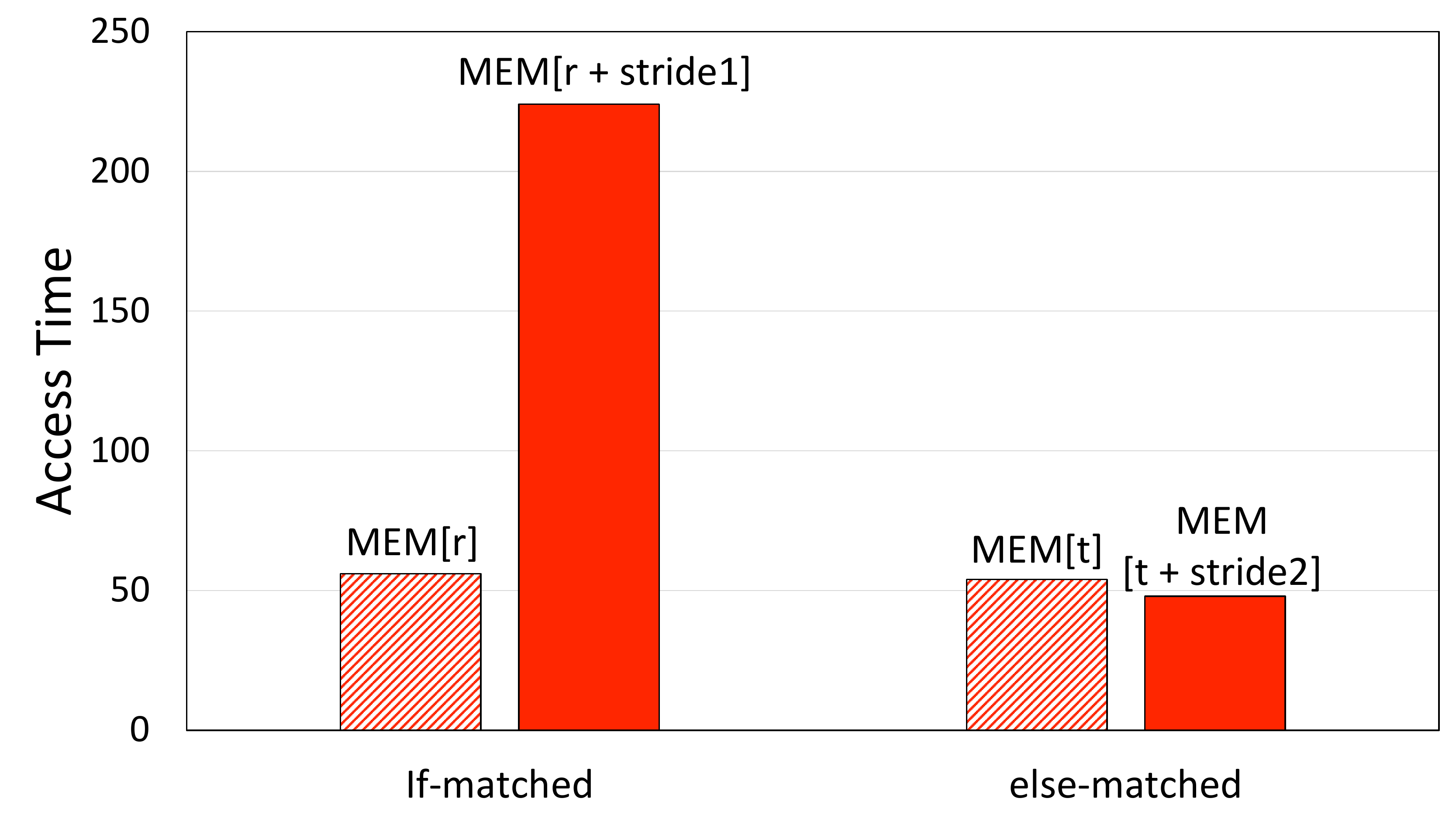}
    \caption{Detecting control flow in \name{} variant 1 without cache probing. The prefetcher can no longer be triggered if the victim executes the target load (and branch). %
    }
    \label{fig:v1_new}
\end{figure}

\textbf{Protected cache.} Other solutions based on randomization~\cite{CEASER, 10.1145/3307650.3322246}, randomize cache indexing, which may hide the stride if the attacker would like to construct the minimal eviction set in the Prime+Probe. \name{}, however, supports using Flush+Reload to observe the cache status variation, which also saves the trouble of computing eviction sets. Other works~\cite{10.1016/j.asoc.2016.09.014, 9251956, zhang2016cloudradar} have been proposed to prevent Flush+Reload by tracking abnormal events (e.g., a large number of fetching, a large number of LLC/L1D cache misses, etc.) and by prohibiting data or instruction fetching. However, the important thing to note is that the attacker can use \name{} in other ways beside Flush+Reload or Prime+Probe. For example, according to our reverse engineering results, after the victim touches a well-trained IP in the prefetcher, the IP's confidence will be updated and it will no longer be able to trigger the prefetcher. The attacker can re-execute all of the well-trained IPs to determine which IPs are no longer triggered and infer the victim's control flow, i.e., checking the \textbf{prefetcher status}. By using this method, instead of reloading or probing the whole page (memory), the attacker only needs to test the latency of a single destination address. To verify the feasibility of the measurement model, we run \name{} variant 1 again. Instead of using Prime+Probe, however, we randomly access two memory addresses using if-matched IP and else-matched IP in the \texttt{gadget}. The experimental result is shown in Figure~\ref{fig:v1_new}. In this example, we find that the if-matched IP no longer triggers the prefetcher, which implies that the victim has executed the if-path.
One limitation of this measurement is the potentially higher false-positive rate. For example, the prefetcher also might not be triggered due to memory accesses at the context switch or because of a large amount of activity. %
This technique, however, allows us to bypass the defense against Flush+Reload or Prime+Probe. %

\textbf{Access control.} There are several works aimed at creating a memory safety zone to prevent unauthorized access to confidential data. Most of them leverage hardware and/or software-based fencing/delaying solutions~\cite{loughlin2021dolma, qi2021spectaint, taram2019context, wang2019oo7} to restrict the speculative loading of secret-related data. More specific, these defenses will selectively serializing the speculative  \texttt{load} instructions by adding \texttt{lfence}-type instructions or directly delaying the speculative \texttt{loads} in the issue queue. These defenses are mainly effective in prohibiting the unauthorized access of secret data in case of transient execution attacks. \name{}, however, doesn't rely on branch-speculative execution to leak the secret. In addition, based on our observations, serializing load instructions will not prevent the prefetcher from generating prefetch requests. %

\subsection{Mitigation Options}

There are some generic defenses that could be deployed to mitigate all variants of \name{}. The most straightforward protection is to disable the IP-stride prefetcher. This defense, however, will result in a significant performance overhead. If possible, redesigning the application to avoid speculative accesses of secrets can also prevent this issue~\cite{intel-guidelines}, if the source of the application is available. Further, oblivious execution~\cite{190908, yu2019data} could prevent data leakage by removing any control flow and most data dependencies. Nevertheless, the use of oblivious code faces practical difficulties, as it leads to significant overhead in many applications and may not work correctly due to the limitations of application or programmer experience~\cite{puddu2020frontal}. %
The use of a secure timer that can obfuscate the cache access latency by adding noise~\cite{130768, 6237011} is another way to mitigate \name{}.
But they have to be built on specific kernel or extended ISA, which are often costly to implement. %
In addition, caching page table entries of sensitive data in a isolated cache rather than traditional caches (e.g., CATalyst~\cite{7446082}) can also mitigate \name{}. However, having a separate cache for page table pages can introduce a high overhead in hardware~\cite{gras2017aslr}.
Our proposed mitigation strategies can be implemented today, or a low-overhead version can be implemented in future hardware designs (See Section~\ref{sec:leak-mitigation}).

\section{Related Work}
\label{sec:related}

Microarchitectural side-channel attacks exploit %
shared resources in the processor to leak sensitive information across privilege domains~\cite{ge2018survey,osvik2006cache}. Examples such as caches, the branch predictor unit (BPU) and 
prefetchers have been successfully demonstrated as potential sources of risk.

\subsection{Cache Side-Channels}

Cache side-channels, e.g., Evict+Time~\cite{osvik2006cache}, Flush+Flush~\cite{gruss2016flush+}, Flush+Reload~\cite{yarom2014flush+}, Prime+Probe~\cite{osvik2006cache,percival2005cache}, utilize the timing variance caused by different memory behaviors. The large latency difference between a cache hit and a cache miss allows for %
high resolution, %
low noise %
observations. %
In addition, the hierarchical structure of the cache allows for information to be leaked %
across different shared timing level, from private data cache and instruction cache, to the last level cache shared by all the cores~\cite{icache,osvik2006cache,yarom2014flush+}.
In many works, secret-related activities can be exhibited on the timing of the cache by nature or by induction~\cite{puddu2020frontal,ren2021see}. Therefore, cache timing is often the basis of manifesting new side-channels. 

\subsection{Prefetching Side-Channels}

The work of Shin et al.~\cite{ccs} is the closest work with ours in leaking information through the IP-stride prefetcher on top of cache side-channels. %
They observe that, when the program itself shows a stable memory access behavior, e.g. table look-ups, the IP-stride prefetcher can be triggered and leave special footprint in the cache. Based on this insight, they are able to leak the secret value of one specific algorithm of the OpenSSL library~\cite{OpenSSL}, the ECDH algorithm~\cite{ECDH}.
Instead, in this work, we present a generic methodology which can extract information from a variety of workloads using the IP-stride prefetcher. %
Detecting only the selected strides in the cache saves huge amount of analyzing effort for the observer. For example, this state-of-the-art work~\cite{ccs} can only find two cache lines that is related to the secret, so that they sample the activities of these two cache lines into time-series data and then adopt clustering algorithm to analyze their correlation with the secret. 

From the software perspective, a software prefetch side-channel~\cite{gruss2016prefetch} aims to bypass Supervisor Mode Access Prevention (SMAP) and KASLR. To infer the mapping between virtual address and physical address in the system, they exploits \texttt{prefetch} instructions' capability of leaking timing information on the exact translation level of the virtual address. %
However, the main purpose of that work is different from \name{}. Moreover, they implement their attack using software interface, while we directly manipulate hardware prefetcher.%

In many previous security-related works, prefetching is regarded as
a hindrance, limiting the ability of the attacker to interpret the results of the victim.
The work that introduces Flush+Reload~\cite{yarom2014flush+} mentions that results from data prefetching should be identified and filtered out when analyzing the results. Tromer et al.~\cite{tromer2010efficient} propose the use of pointer-based traversal method in Prime+Probe to suppress prefetching. 
Wang et al.~\cite{papp} eliminate prefetching effect from Prime+Probe by constructing eviction sets with an extra warm-up section and designs special priming and probing sequences on the specific machines. Some works \cite{fang2018prefetch, fuchs2015disruptive} even exploit the prefetcher as a defence against a covert channel and cache side-channel by obfuscating the communication synchronization or cache footprint itself.

\subsection{Branch Predictor-based Side-Channels}
The branch predictor (BPU, branch prediction unit), another shared structure, has been exploited to conduct speculative execution based attacks and has been studied extensively~\cite{bhattacharyya2020specrop,Kocher2018spectre,lipp2018meltdown,ren2021see,schwarz2019zombieload,van2020cacheout}. In these works, the branch predictor can be mis-trained by the adversary to force speculative execution of mispredicted instructions. We find that the microarchitecture of the Intel BPU resembles its hardware prefetcher in many aspects. For example, they are shared among all applications operating on the same core, and they both have confidence counter mechanisms. Kocher et al.~\cite{Kocher2018spectre} proposed \textit{Spectre} to exploit conditional branch misprediction and poison indirect branches. Although the mispredicted branch will not be committed in the end, system information, such as cached data, can be affected. Our implementation is similar to Spectre's in mistraining a hardware component in the processor and exploiting cache timing to analyze results. Nevertheless, compared to the BPU, the hardware prefetcher will not introduce additional instructions for execution or cause pipeline flush events; it only brings load data into the cache.
The IP-stride prefetcher takes a relatively short amount of time to %
train compared to the BPU. Also, since BPU 
uses the least 20 bits of IP, Spectre still needs several rounds of testing to bypass ASLR, while the IP-stride prefetcher uses the least 8 bits and is not affected.

\subsection{Other Microarchitectural Side-Channels}
Apart from the traditional cache, BPU and prefetcher, some other architectural components have recently received attention as well. Ren et al.~\cite{ren2021see} reveals the characteristics of the micro-op cache in Intel and AMD processors and exploits them as a timing channel to transmit secret information. The TLB, or Translation Look-aside Buffer, is another shared resources in the processor, can also leak information. An OS-level adversary can induce page faults to observe the page-level access patterns of the victim~\cite{wang2017leaky}. TLBleed~\cite{gras2018translation} further combines machine learning methods that improves its granularity. CacheOut~\cite{van2020cacheout} details how an attacker can leak information cross multiple security boundaries by using the line-fill buffer (LFB).
Recently, a front-end side-channel is proposed~\cite{puddu2020frontal}. They found that during an interrupt, some instructions’ execution time will change when instructions around them, and their virtual addresses varies. They managed to exploit this timing difference on Intel SGX to distinguish between instruction-wise identical branches and extract secret state. %

\section{Conclusion}
\label{sec:conclusion}
In this work, we observe that Intel IP-stride prefetcher is a shared hardware resource. We leverage this feature to introduce a novel side-channel attack that can leak control flow named \name{}. To accomplish the attack, we present an in-depth study of the Intel IP-stride prefetcher, revealing %
a number of
undocumented details. \name{} is able to leak victim's control flow cross (a) code regions, (b) process spaces, and (c) cross user-kernel boundary. We show that \name{} achieves a success rate of up to 99\%, depending on variant. 
We finally discuss effectiveness of existed defenses and propose two mitigation techniques to block this side-channel, one of which can be used on hardware systems today.

\bibliographystyle{plain}
\bibliography{ref}
\justifying

\end{document}